\documentclass[aps,prd,twocolumn,amsmath,fleqn,superscriptaddress,showpacs]{
revtex4-1}

\usepackage{amssymb}

\usepackage[pdftex]{graphicx}

\def\apjl{ApJL} 
\def\aj{AJ} 
\def\apj{ApJ} 
\def\pasp{PASP} 
\def\aap{A\&A} 
\def\nar{New Astron. Rev.} 
\def\mnras{MNRAS} 
\def\jcap{JCAP} 

\usepackage{hyperref}  

\newcommand{\adsurl}[1]{\href{#1}{ADS}} 
\providecommand{\url}[1]{\href{#1}{#1}}

\usepackage{sidecap}
\usepackage{subfigure}
\usepackage[subfigure]{ccaption}

\newcommand\fig[1]{Figure~(\ref{#1})}

\newcommand\eq[1]{Eq.~(\ref{#1})}
\newcommand\eqs[2]{Eqs.~(\ref{#1}) and (\ref{#2})}

\newcommand\eqsx[6]{Eqs.~(\ref{#1}), (\ref{#2}), (\ref{#3}), 
(\ref{#4}), (\ref{#5}) and (\ref{#6})}

\newcommand{\be}{\begin{equation}}
\newcommand{\ee}{\end{equation}}

\newcommand{\bx}{{\bf x}}
\newcommand{\bxi}{{\bf x}_{\bf i}}
\newcommand{\bxj}{{\bf x}_{\bf j}}
\newcommand{\hbx}{\hat{{\bf x}}}

\newcommand{\bv}{{\bf v}}
\newcommand{\bk}{{\bf k}}
\newcommand{\bC}{{\bf C}}

\begin{document}


\title{Data compression of measurements of peculiar velocities of
  Supernovae Ia} 

\author{Vid Ir\v{s}i\v{c}} 
\affiliation{Faculty of Mathematics and Physics, 
 University of Ljubljana, Jadranska 19, 1000 Ljubljana, Slovenia}

\author{An\v{z}e Slosar}
\affiliation{Brookhaven National Laboratory, Upton NY 11973, USA}

\begin{abstract}
	We study the compression of information present in the correlated
	perturbations to the luminosity distance in the low-redshift
	($z<0.1$) supernovae Ia due to peculiar velocities of these
	supernovae. We demonstrate that the na\"{i}ve compression into
	angular velocity power spectrum does not work efficiently, due to
	thickness of the spherical shell over which the supernovae are
	measured. Instead, we show that measurements can be compressed into
	measurements of $f^2P(k)$, where $f$ is the logarithmic rate of
	growth of linear perturbations and $P(k)$ is their power spectrum.
	We develop an optimal quadratic estimator and show that it recovers
	all information for $\Lambda CDM$ models for surveys of
	$N\sim10,000$ or more supernovae. We explicitly demonstrate
	robustness with respect to the assumed fiducial model and the number
	of power spectrum bins. Using mock catalogues of SNe Ia we estimate 
	that future low redshift surveys will be able to probe $\sigma_8$ to 
	$6\%$ accuracy with $10,000$ SNe Ia.
\end{abstract}


\maketitle

\section{Introduction}
\label{sec:introduction}
Since the earliest studies of supernovae, it has been suggested that
type Ia (SNe Ia) might be used as standard candles for cosmological
measurements. In the subsequent years many studies using SNe Ia
revealed that the expansion of the universe is accelerating
\cite{1998AJ....116.1009R,1999ApJ...517..565P}. From SNe Ia
observational data one could measure cosmological parameters
describing the homogeneous expansion of the universe through
measurements of the luminosity distance such as matter and dark energy
densities and equation of state parameters
\cite{2006A&A...447...31A,2007ApJ...659...98R,2007ApJ...666..694W,
2002SPIE.4836.....T,2005AAS...20718005D,2006PASP..118....2H,2005NewAR..49..346A}
.

But density inhomogeneities cause additional scatter to SNe Hubble
flow
\cite{1995ApJ...445L..91R,1997ApJ...488L...1R,1998ApJ...503..483Z,
2000AAS...196.6207B,2004MNRAS.355.1378R,2006PhRvL..96s1302B,2007ApJ...661..650H,
2007ApJ...659..122J,2007MNRAS.379..343W,2007ApJ...664L..13C,2007arXiv0705.0368W,
2007ApJ...661L.123N,2008JCAP...02..022H,1988ApJ...332L...7G}.
In the recent years it has become apparent that the correlations
between these peculiar velocities of the supernovae are significant at
low redshifts ($z < 0.1$)
\cite{2006PhRvD..73l3526H,2007PhRvL..99h1301G} and thus present an
opportunity to measure cosmological parameters that affect the growth
of perturbations in the Universe such as amplitude and shape of the
matter power spectrum.

But extracting any additional information from the correlations
induced by peculiar-velocities comes at an expense of computing
$N_{sn} \times N_{sn}$ correlation matrix and its inverse on every
likelihood evaluation. To complicate matters even further,
each matrix element consists of an oscillatory integrals thus adding
to the cost of the overall computing time.  In this paper we asses
whether a data compression method can be found that will provide both
efficiency and simplified form of the cosmological information carried
by the peculiar velocity correlation matrix. This will also provide a
physical insight into what is that the peculiar velocities of
supernovae Ia are measuring.

Through the paper we assume the correlated part of peculiar velocities
is described by the linear theory and that small scale virial
velocities can be described by single parameter describing the
velocity dispersion. While this is probably not a good approximation
for realistic surveys of thousands of supernovae (after all,
supernovae don't measure velocities at random positions and the real
Universe is not linear), it nevertheless provides a good simple
framework for studying an efficient information compression.

The paper is structured as follows. In Section \ref{intro-corr} we
discuss the background theory required to describe correlations in the
luminosity distances to nearby supernovae sourced by correlated
large-scale velocities.  In Section \ref{sec:proj-veloc-angul} we
discuss the angular power spectrum of the projected velocities as a
candidate for data compression. In Section \ref{Pksec} we discuss the
power spectrum multiplied by a factor describing the growth
of fluctuations at mean redshift as our proposed method for data
compression.  We conclude in Section \ref{conclusion}.

Throughout the paper we assume that unless noted otherwise, the
variables have their usual meaning.


\section{Correlations of peculiar velocities}\label{intro-corr}

The luminosity distance $d_L$ to a distant SN at redshift $z$ is
defined by \be F = \frac{L}{4\pi d_L^2}, \label{flux} \ee where $F$ is
the measured flux and $L$ the intrinsic luminosity of the
supernovae. In the Friedmann-Robertson-Walker (FRW) metric, the
luminosity distance is related to the comoving distance to an object at
redshift $z$ as \be d_L(z) = (1+z)\left\{ \begin{array}{lr}
    \frac{1}{\sqrt{k}}\sin(\chi(z)\sqrt{k}) & k > 0 \\
    \chi(z) & k = 0 \\
    \frac{1}{\sqrt{-k}}\sinh(\chi(z)\sqrt{-k}) & k < 0
  \end{array} \right., \label{lumdist}
\ee
where $k$ is the geometrical curvature of the universe and we have introduced
$\chi(z)$ as the comoving distance to a distant object at $z$ as
\be
\chi(z) = \int_0^z \frac{dz'}{H(z')}. \label{comoving-x}
\ee
Astronomers prefer the flux relation (\eq{flux}) rewritten in terms of the
magnitudes as
\be
m - M = 5 \log_{10} \left( \frac{d_L}{\mathrm{Mpc}} \right) + 25,
\ee
where $m$ and $M$ stand for the apparent and absolute magnitude respectively.

The perturbed FRW metric leads to the effect of peculiar velocities
and those in turn lead to the perturbations in luminosity distance
given by
\cite{2006PhRvD..73b3523B,2006PhRvD..73l3526H,2007PhRvL..99h1301G} \be
\frac{\delta d_L}{d_L} = \frac{d_L^{me} - d_L^{th}}{d_L^{th}} = \hbx
\cdot \left( \bv - \frac{(1+z)^2}{H(z)d_L^{th}} (\bv - \bv_o)
\right), \label{distfl} \ee where $d_L^{me}$ and $d_L^{th}$ stand for
measured luminosity distance and theoretical prediction for
unperturbed space-time, given by \eq{lumdist}. Velocities $\bv$ and
$\bv_o$ are peculiar velocities of the source and the observer
respectively. Projection of peculiar velocities along the line of
sight ($\hbx$) is the only component we can measure. Redshift $z$ on
the right-hand side of the equation stands for the observed redshift.

Since the peculiar velocities result from some initial Gaussian matter
perturbations, the peculiar velocity measurements are drawn from a
distribution with zero mean and nonzero variance. The later can be
written in the form of correlation function as $\xi(\bx_1,\bx_2) =
\langle (\bv(\bx_1)\cdot \hbx_1) (\bv(\bx_2)\cdot \hbx_2)^* \rangle$
for two SN at comoving position $x_{1,2}$. The correlation function of
projected peculiar velocities has been computed in a number of studies
\cite{2007PhRvL..99h1301G,Dodelson}. The result, assuming linear
perturbation theory, can be written as
\begin{align}
\xi(\bx_1,\bx_2) =& \sin{\theta_1}\sin{\theta_2}\xi_{\perp}(x,z_1,z_2) + \notag
\\
& + \cos{\theta_1}\cos{\theta_2}\xi_{\parallel}(x,z_1,z_2), \label{corr-func}
\end{align}
where $\bx_{12} \equiv \bx_1 - \bx_2$, $x = |\bx_{12}|$,
$\cos{\theta_1} \equiv \hbx_1 \cdot \hbx_{12}$ and $\cos{\theta_2}
\equiv \hbx_2 \cdot \hbx_{12}$. The projections of the correlation
function are given by \cite{2007PhRvL..99h1301G,Dodelson} \be
\xi_{\parallel,\perp} = {\bar D}'(z_1) {\bar D}'(z_2)
\int_{0}^{\infty}
\frac{dk}{2\pi^2}P(k)K_{\parallel,\perp}(kx),\label{xiperp} \ee where
${\bar D}(z)$ is the normalized growth function (${\bar D}(z) =
D(z)/D(z=0)$) and derivatives are with respect to conformal time ($' =
d/d\eta$, $a(t)d\eta = cdt$). $P(k)$ is the matter power spectrum and
integration kernels are given by $K_{\parallel}(x) = j_0(x) -
\frac{2j_1(x)}{x}$ and $K_{\perp}(x) = \frac{j_1(x)}{x}$.

Correlations in the peculiar velocity lead to correlations in the
luminosity distance fluctuations which can be written as
\cite{2007PhRvL..99h1301G} \be C_v(i,j) = \left(1 -
  \frac{(1+z)^2}{H(z)d_L}\right)_i \left(1 -
  \frac{(1+z)^2}{H(z)d_L}\right)_j \xi(\bxi,\bxj).
\label{cvij}
\ee The total correlation matrix is a sum of peculiar velocity
correlation matrix and uncorrelated scatter $\sigma_i^2$ and can be
written as \cite{2007PhRvL..99h1301G} \be C_{ij} = C_v(i,j) +
\delta_{ij} \sigma_i^2.
\label{cij}
\ee The diagonal parts of the matrix given by
\cite{2007PhRvL..99h1301G} \be \sigma_i^2 =
\left(\frac{\ln{10}}{5}\right)^2 (\sigma_m^2 + \sigma_{m_i}^2) +
\left(1 - \frac{(1+z)^2}{H(z)d_L}\right)_i^2
\sigma_v^2, \label{uncorr} \ee where $\sigma_{m_i}$ stands for
observational errors on apparent magnitudes, which are a property of
the dataset and vary from supernova to supernova. The remaining two
parameters are $\sigma_m$, which is intrinsic magnitude scatter
describing deviations of supernova luminosities from the perfect
standard candles, while $\sigma_v$ is the small scale velocity
dispersion due to uncorrelated small scale peculiar velocities.


\section{Projected velocity angular power spectrum}\label{angular}
\label{sec:proj-veloc-angul}

Measurements of the SNe are most commonly transformed into luminosity
distances or, in case of peculiar velocities, into luminosity distance
fluctuations. But there is no apparent reason why not to work with the
projected peculiar velocity field instead
\cite{2008JCAP...02..022H}. Since we are interested in the velocity
perturbations along the line of sight, the projected peculiar velocity
field seems more natural as well. Those velocity perturbations can be
understood as perturbations relative to the expansion of the universe
in physical coordinates, or as perturbations with respect to the
comoving grid.

Since the peculiar velocity field is a scalar function on a sphere, it
is natural to consider its angular power spectrum. Using expansion over
spherical
harmonics we get \be \bv(z) \cdot \hat{r}z =\sum_{\ell=0}^{\infty}
\sum_{m=-\ell}^{+\ell} a_{\ell m}(z) Y_{\ell
  m}(\theta,\phi), \label{PVfield} \ee where we kept the dependence of
the expansion on redshift $z$, i.e. we consider infinitely thin
spherical shells at redshift $z$. $Y_{\ell m}$ are
spherical harmonic functions that satisfy orthonormal relation \be
\int d\Omega Y_{\ell m}(\theta,\phi) Y^*_{\ell'm'}(\theta,\phi) =
\delta_{\ell \ell'} \delta_{mm'}. \label{ortonormal} \ee Using the
assumption of Gaussian initial perturbations we can relate angular
auto and cross power spectra to the expansion coefficients ($a_{\ell m}$) as \be
\langle a_{\ell m}(z) a^*_{\ell'm'}(z') \rangle = C_\ell(z,z') \delta_{\ell
\ell'}
\delta_{mm'}. \label{alm-def} \ee

In the linear theory, these are given by 
\begin{align}
C_\ell(z, z') &= \bar{D}'(z) \bar{D}'(z') \frac{2}{\pi} \cdot \notag \\
& \cdot \int_0^\infty dk P(k) \left( \frac{\partial
    j_\ell(k\chi)}{\chi \partial k} \right) \left( \frac{\partial
j_\ell(k\chi')}{\chi'\partial k} \right),
\label{Cls}
\end{align}
for two SNe at the redshifts $z$ and $z'$, where $\bar{D}$ is the
growth factor, $j_\ell$ are spherical Bessel functions, $\chi^{(')} =
\chi(z^{(')})$ and $P(k)$ matter power spectrum. The detailed
derivation can be found in the Appendix.

The angular power spectrum of luminosity distance fluctutations are
exactly the same as in \eq{Cls} save for the prefactors and can be
written as
\begin{align}
D_\ell(z,z') &= \left(1 -
  \frac{(1+z)^2}{H(z)d_L(z)}\right) \cdot \notag \\
& \cdot \left(1 -
  \frac{(1+z')^2}{H(z')d_L(z')}\right) C_\ell(z,z').
\label{Cvs}
\end{align}

Both angular power spectra have dimensionless units. This is due to the fact
that velocity in our all calculations is in the units of the speed of light.

\begin{figure}
\centerline{\includegraphics[width=8.0cm]{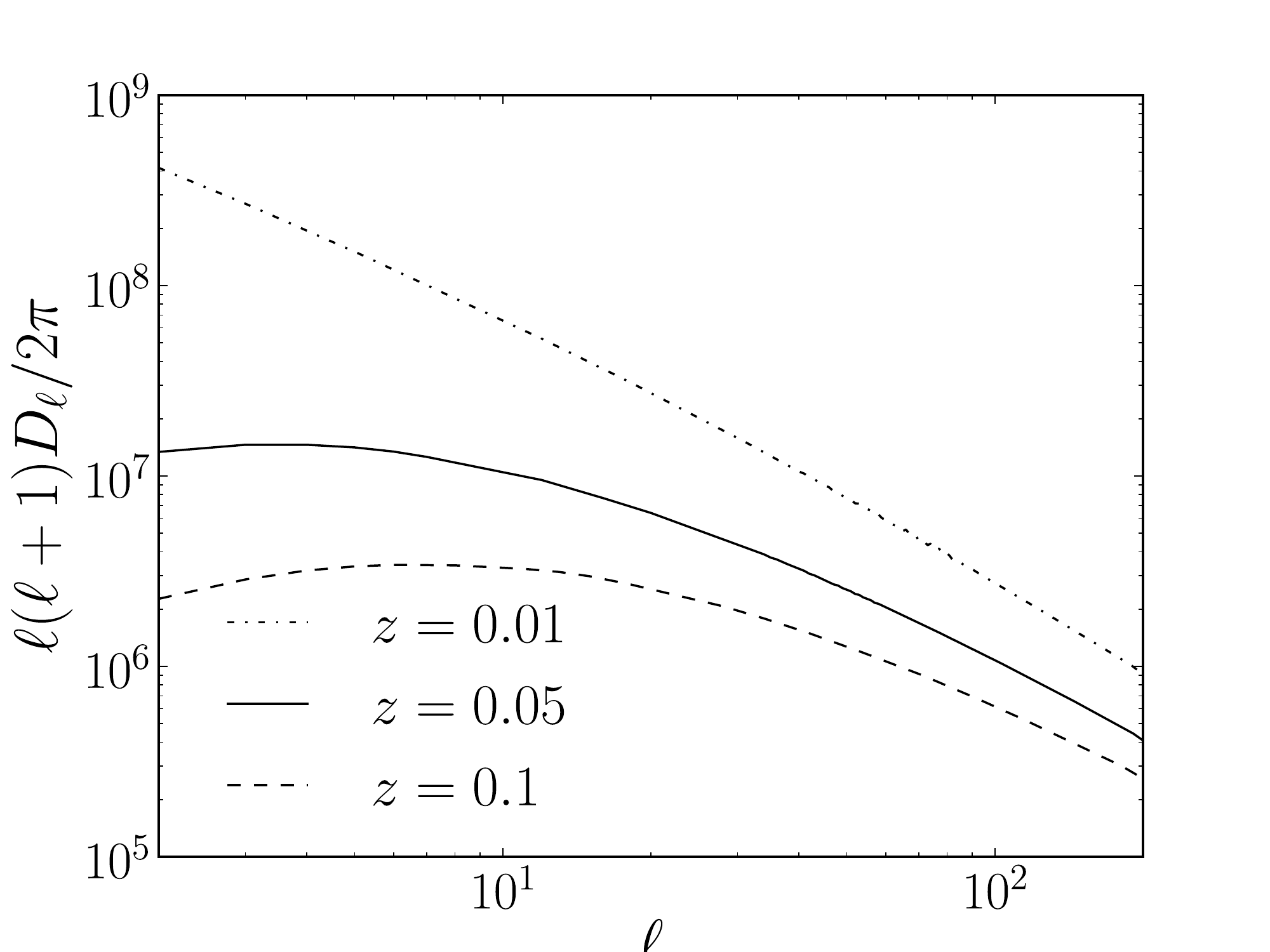}}
\caption{The luminosity distance fluctutations angular power spectrum plotted as
  $\ell(\ell+1)D_\ell/2\pi$ vs $\ell$ for three different redshifts:
  $z=0.05$ (solid), $z=0.1$ (dashed) and $z=0.01$ (dot-dashed). The angular
power spectrum has dimensionless units.  }
\label{cls-redshift}
\end{figure}

\fig{cls-redshift} shows the auto-power spectrum $D_\ell (z,z)$ for
different values of redshift. As expected and also shown in the
literature \cite{2008JCAP...02..022H}, numerical simulations are in
good agreement with above result (\eq{Cls}) only for small $\ell$
($\ell < 100$), where linear regime is still valid.

\begin{figure}
\centerline{\includegraphics[width=8.0cm]{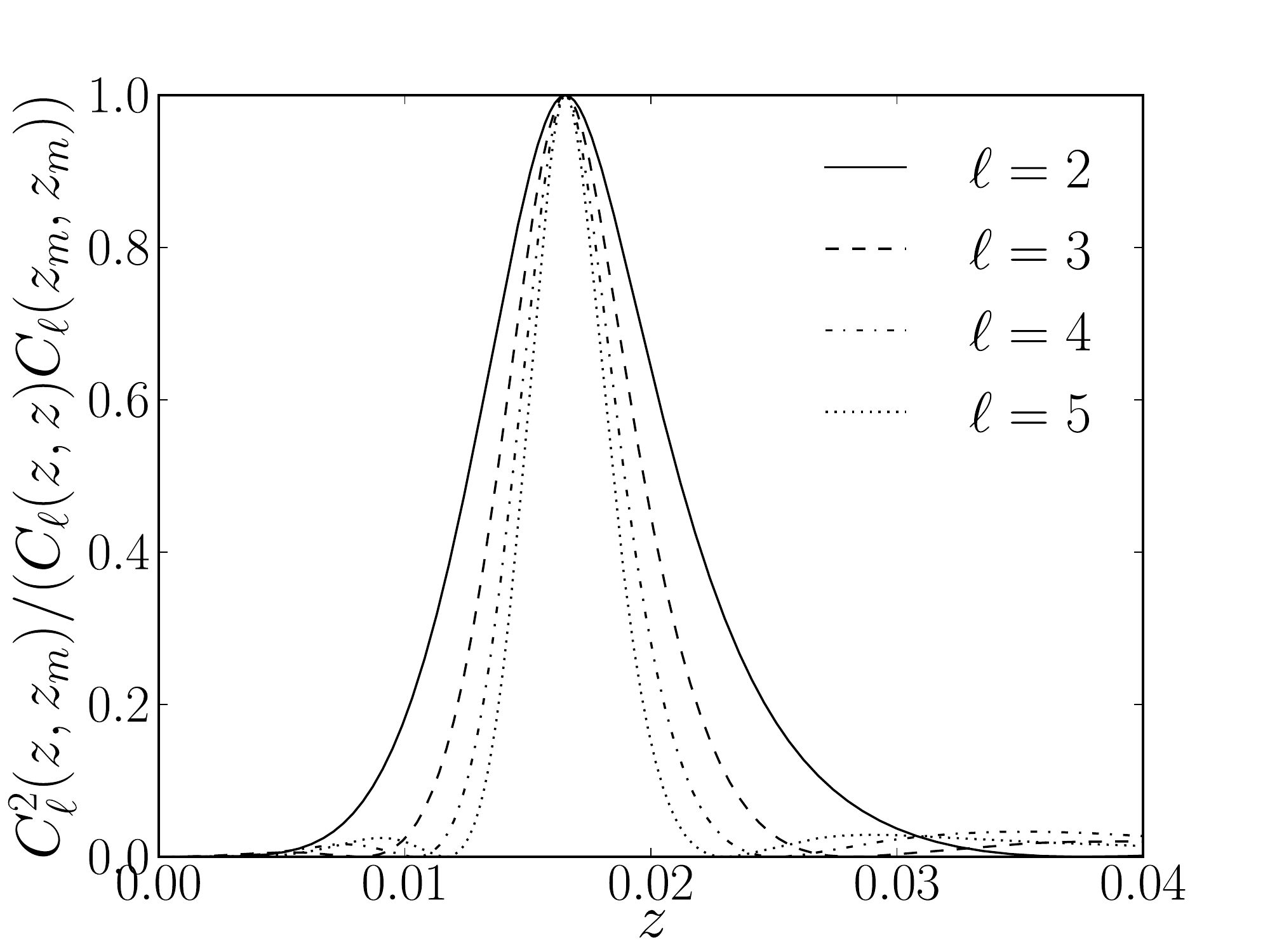}}
\caption{The cross correlation coefficient $r(z)$ (Eq. \ref{eq-corr})
  with respect to reference redshift $z_{m} = 0.016528$. The
  coefficient is plotted for four low multipoles: $\ell=2$ (solid),
  $\ell=3$ (dashed), $\ell=4$ (dot-dashed) and $\ell=5$ (dotted).  }
\label{cls-corr}
\end{figure}

Although the formalism is appealing, is the angular power spectrum
actually a useful compression method for the low-redshift supernovae?
If we want to describe all peculiar velocities of supernovae at low
redshifts using a single power spectrum, the slices at different
redshifts would essentially need to be describing the same velocity
field. In order to check this assumption, we look at the
cross-correlation coefficient given by \be r (z,z') =
\frac{C_\ell^2(z,z')}{C_\ell(z,z)C_\ell(z',z')}.
\label{eq-corr}
\ee 

In \fig{cls-corr} we plot $r(z,z')$ while fixing $z'=0.016$, a
median redshift of a typical low-z SNe dataset used in
\cite{2007ApJ...659..122J}. We see that the 'correlation length' of
velocities in redshift is only about $\delta z= 0.01$ and hence to
describe all correlations between $z=0.0$ and $z=0.1$, we would need
about 10 angular power spectra, together with a large covariance
matrix that will describe not only the errors and their covariances at
one redshift, but also those at neighbouring redshifts. This makes
this approach clearly suboptimal and hence we turn the direct
estimation of the three-dimensional power spectrum.

\section{Effective power spectrum measurements}
\label{Pksec}

In any data compression technique, we try to put constraints on
quantities that are as close to the data as possible and as
independent of the underlying theoretical assumptions, while at the
same time still capturing most of the information in the full
dataset. Although the covariance matrix of perturbations to the
luminosity distance depends on a number of cosmological dependent
pre-factors, in addition to the underlying power spectrum $P(k)$, we
show that, in the limit of supernovae being at sufficiently low
redshifts, the data effectively measure $P(k)f^2$, where $f$ is the
logarithmic growth rate, regardless of which fiducial cosmology one
adopts.

Since the $P(k)$ is a continuous function of the wave vector $k$ we
will approximate it with a stepwise function such that $P(k) =
\text{const.}$ for $ k_\alpha \leq k < k_{\alpha+1}$. With this
in mind we can rewrite the parallel in perpendicular projections of
correlation function of two SNe ($i$ and $j$) as
\begin{align}
\xi_{\parallel}([k_\alpha,k_{\alpha+1})) =& \bar{D}'(z_i) \bar{D}'(z_j)
\frac{P(k_\alpha)}{2\pi^2 x} \int_{k_\alpha x}^{k_{\alpha+1} x} K_\parallel(y)
\, dy \notag \\
=& \bar{D}'(z_i) \bar{D}'(z_j) \frac{P(k_\alpha)}{2\pi^2 x} \cdot \notag \\
&\cdot \left( j_1(k_{\alpha+1} x) - j_1(k_\alpha x) \right),
\label{corr-par-stepwise}
\end{align}

\begin{align}
\xi_{\perp}([k_\alpha,k_{\alpha+1})) =& \bar{D}'(z_i) \bar{D}'(z_j)
\frac{P(k_\alpha)}{2\pi^2 x} \int_{k_\alpha x}^{k_{\alpha+1} x} K_\perp(y) \, dy
\notag \\
&= \bar{D}'(z_i) \bar{D}'(z_j) \frac{P(k_\alpha)}{4\pi^2 x} \cdot \notag \\ 
&\cdot \left( \text{Si}(k_{\alpha+1} x) - \text{Si}(k_\alpha x)\right) -
2\xi_{\parallel},
\label{corr-perp-stepwise}
\end{align}
where $x = |\bx_{ij}|$, $\bx_{ij} = \bx_i - \bx_j$ and $\bx_i =
\chi(z_i)\hbx_i$ (\eq{comoving-x}), $\bar{D}(z)$ is normalized growth
function and $j_1$ are spherical Bessel functions of the first
kind. We have also defined integral sinus as $\text{Si}(x) = \int_0^x
\frac{\sin{z}}{z}dz$. \eqs{corr-par-stepwise}{corr-perp-stepwise} are
valid only when $x \neq 0$. Because $x$ is the norm of the difference
of two vectors to SN we will surely have examples where $x=0$
(peculiar velocity auto-correlation functions). When taking $x
\rightarrow 0$ the above expressions simplify to \be
\xi_{\parallel,\perp}([k_1,k_2)) = \bar{D}'(z_i) \bar{D}'(z_j)
\frac{P(k_1)}{6\pi^2} \left( k_2 - k_1 \right).
\label{corr_x0_stepwise}
\ee

Parameters we want to constrain from the SNe Ia data are values of
matter power spectrum $P_\alpha$, and two uncorrelated errors
$\sigma_m$ and $\sigma_v$ and so $N_p = N_b + 2$. We have assumed that
the absolute magnitude offset will be for all our purposes completely
constrained from all supernovae data (including higher-redshift ones).


\subsection{Mock catalogues}\label{mock}
\vspace{-0.4cm} To test our method we created several mock catalogues
of SNe redshifts and their positions on the sky. For those catalogues
we calculated their synthetic observational data in the form of
luminosity distance fluctuations. We chose to model are synthetic data
with fiducial cosmological parameters $(\Omega_m = 0.24, h = 0.7, w =
-1, \Omega_\Lambda = 1 - \Omega_m)$. We computed the covariance matrix
give by \eq{cij} using reference linear matter power spectrum
$P^{ref}(k)$ computed using CAMB\footnote{Matter power spectrum was
  computed with CAMB using the following cosmological parameters
  ($\Omega_m = 0.24,\Omega_b = 0.04,h=0.7,w = -1, n_s = 1, \sigma_8 =
  0.789347$)}. We computed eigenvalues $\lambda_i$ and eigenvectors
$v_{\lambda_i}$ of covariance matrix and than, for each $\lambda_i$,
draw a random number from a Gaussian distribution with
$\sigma=\sqrt{\lambda_i}$ and add this number to the initially zero
data vector in the direction of the $v_{\lambda_i}$. This procedure
gave us at the end a data vector of luminosity distance fluctuations
that was a linear combination of every eigenvector.

Uncorrelated errors of apparent magnitudes were randomly chosen from a
uniform distribution between $0.05$ and $0.2$ mag. Moreover we assumed
that any errors in the redshift are negligible. Uncorrelated magnitude
and peculiar velocity scatter had fixed values of $\sigma_m =
0.1\;\mathrm{mag}$ and $\sigma_v = 300\;\mathrm{km/s}$.


\subsection{Optimal quadratic estimator}\label{oqe}

Because of the simplified form of our correlation matrix we chose the
Newton iteration method for zero-finding of the derivative of
probability function $L$, given by \cite{Dodelson} \be \ln{L} = -
\frac{\Delta^T \bC^{-1} \Delta}{2} - \frac{1}{2}\ln
\left(\mathrm{det}\,\bC\right) - \frac{N_{sn}}{2}\ln{2\pi},
\label{lnL}
\ee where is $\Delta$ the data vector and $\bC$ the correlation matrix
given by \eq{cij}. Our model depends on $a_p$, $p=1,\ldots,N_p$
parameters and is thus multidimensional. The correction to the
parameters $a_p$ for Newton's method is then given by
\cite{Dodelson,1998PhRvD..57.2117B} \be \delta a_p = - \sum_{p'}
\left[ \frac{\partial^2 \ln{L(a)}}{\partial a_p \partial a_{p'}}
\right]^{-1} \frac{\partial \ln{L(a)}}{\partial a_{p'}}.
\label{gauss_est}
\ee The common simplification is to replace the second derivative of
$\ln{L}$ with its ensemble average \cite{Dodelson,1998PhRvD..57.2117B}
\begin{align}
F_{pp'} &\equiv - \langle \frac{\partial^2 \ln{L(a)}}{\partial a_p \partial
a_{p'}} \rangle \notag \\
&= \frac{1}{2}\mathrm{Tr}\left( \bC^{-1}\, \bC_{,p} \,\bC^{-1}\,\bC_{,p'}
\right),
\label{fisherM}
\end{align}
known as the Fisher matrix. When taking the ensemble average we assume
that the underlying theory is correct and that the following relation
is true $\langle \Delta \, \Delta^T \rangle = \bC$.

The correction to the parameter $a_p$ can then be written as
\cite{Dodelson,1998PhRvD..57.2117B} \be \delta a_p =
\frac{1}{2}\sum_{p'} (F^{-1})_{pp'} \,\mathrm{Tr}\left[ \left(
    \Delta\, \Delta^T - \bC \right)\left( \bC^{-1}\, \bC_{,p'}
    \,\bC^{-1} \right) \right].
\label{apoqe}
\ee This method is called the optimal quadratic estimator (OQE) and is
an iterative method. Although the OQE uses the Fisher matrix instead
of the matrix of second derivatives, it converges to the same
maximum. This is true because both matrices ($F$ and the matrix of
second derivatives) are invertible and for both we are in maximum when
$\delta a_p = 0$. The only approximation comes in using the Fisher
matrix to approximate the errors.

Since our parameters ($P_\alpha,\sigma_m,\sigma_v$) have physical
meaning only when they are positive, we have checked this condition on
every iteration step. Were they negative their values were put to
zero.


\subsection{Application to synthetic data}\label{results}

Equation \eq{xiperp} tells that aside from the matter power spectrum,
the peculiar velocity correlation function depends on the cosmology
through the linear growth factor as well. Because we are interested in
low-z SNe ($z<0.1$), where the peculiar velocity effect is still
large, we are in the regime where the Hubble rate is almost constant,
and equal the Hubble constant. In the limit of low-z we can rewrite
luminosity distances for flat universe as \be d_L \approx (1+z)
\frac{c z}{H_0}, \qquad H(t) \approx H_0. \label{lowz-approx} \ee With
this in mind the factors in the luminosity distance correlation matrix
(\eq{cvij}) become independent of cosmological parameters. A little
more care must be exercised with the derivative of the growth
factor. If we expand it into a more suitable form

\begin{align} {\bar D}'(z) = \frac{d{\bar D}(z)}{d\eta} =&
\frac{1}{c}f(a)\,a\,H(a)\,\frac{D(z)}{D(z=0)} \notag \\ =&
\frac{1}{c}f(a)\,\frac{1}{1+z}\,H(a)\,{\bar D}(z),
\label{Dbar}
\end{align} where we have defined logarithmic growth rate $f$ as \be f
\equiv \frac{a}{D(a)}\frac{dD(a)}{da} = \frac{d\ln{D}}{d\ln{a}}.  \ee

Therefore, in this limit we can approximate
\begin{multline}
  {\bar D}'(z) {\bar D}'(z') P(k) \\ \approx \frac{H_0^2}{(1+z)(1+z')} f(z)
  f(z') ({\bar D}(z){\bar D}(z') P(k))\\ \approx
  \frac{H_0^2}{(1+z)(1+z')} \left(f^2({\bar{z}}) P(k,\bar{z}) \right)
\label{Dbar_approx}
\end{multline}

This illustrates that in the limit of small redshifts, the relevant
quantity that our method is sensitive to is a non-dimensional growth
factor multiplying matter power spectrum at some effective redshift
(which we show to be the mean redshift). Our final result for the
correlation matrix is thus algebraic combination of 
\eqsx{cvij}{cij}{corr-par-stepwise}{corr-perp-stepwise}{corr_x0_stepwise}{Dbar_approx}.

\begin{figure}
\centerline{\includegraphics[width=8.0cm]{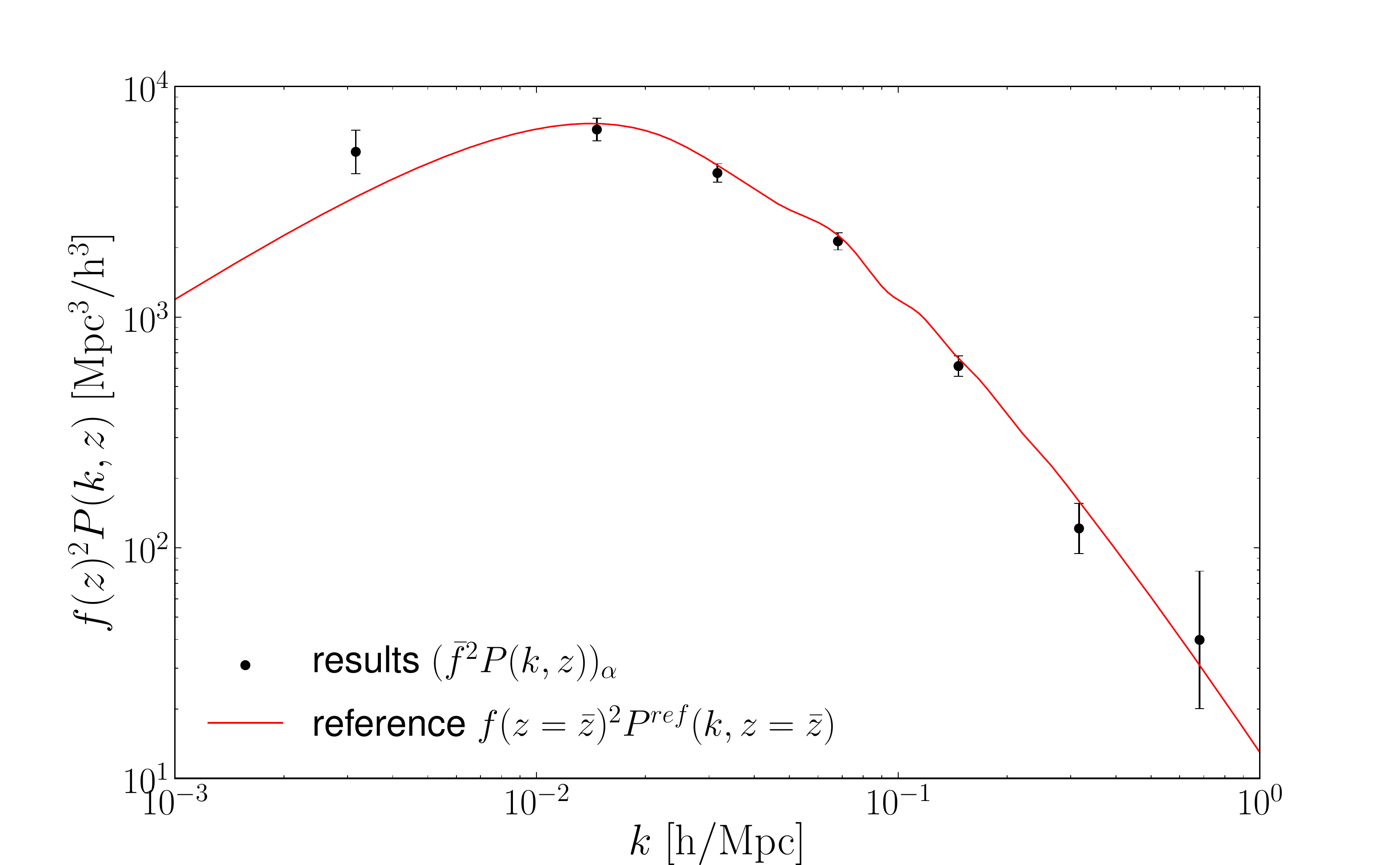}}
\caption{ Shows the result of the new method in the form $(f^2
  P(k,z))_\alpha$ (datapoints with errors) and the underlying model
  that went into creation of synthetic dataset (solid line). \label{fig:nfig} }
\label{PkSNe}
\end{figure}

First, we have tested our method assuming a future survey of 10,000
supernovae Ia. We plot the results, together with the theoretical
model used to create the dataset in \fig{PkSNe}. This illustrates
that the method basically works.  We will proceed with a series of tests.

First, we have tested our method using synthetic data with different
number of SN (1000,5000,10000) and different number of bins $N_b$ on
which we estimated the matter power spectrum. We chose the bin
positions by trial and error, so that the edge bin positions had very
large errors and hence contained very little information. We therefore
converged on the following set-up: one bin on small scales $1-10
h/{\rm Mpc}$, two on large scales $10^{-4} - 10^{-3} h/{\rm Mpc}$ and
$10^{-3}-10^{-2} h/{\rm Mpc}$. The rest were uniformly distributed in
the logarithmic scale  over the interval $0.01 - 1 h/{\rm Mpc}$. We
always discard edge bins which contain essentially no information.

Our method relies on the overall signal-to-noise to be high enough for
the central theorem limit making the power spectrum constraints have
Gaussian errors. To test this, we have performed a simple test. We
analysed the same synthetic data two ways. First we measured the
effective power spectrum from the data and then, for a standard
$\Lambda CDM$ cosmology measured the value of $\sigma_8$, while
keeping all the other cosmological parameters fixed at their fiducial
value. We then repeated the same, but this time skipped the
intermediate test and measured $\sigma_8$ directly from the synthetic
data.  In the limit of infinite number of supernovae, the two should
match perfectly, but we do expect deviations to appear for a small
finite number of datapoints.

\begin{figure}
\centerline{\includegraphics[width=8.0cm]{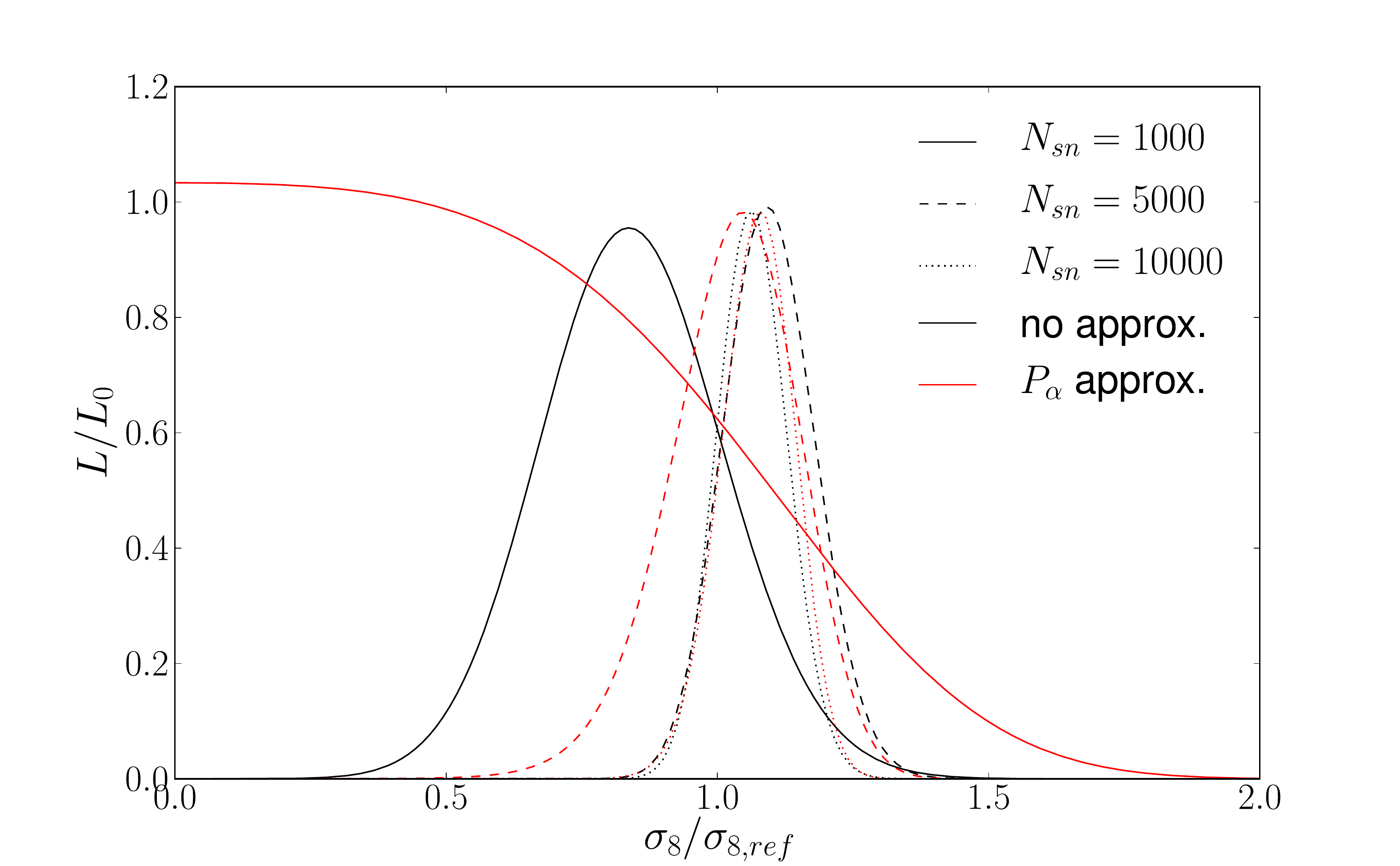}}
\caption{ Show the probability distribution of cosmological parameter
  $\sigma_8$ for data $P_\alpha$ with respect to the reference power
  spectrum. It shows lines representing different number of SNe at the
  constant number of bins $N_b=9$. We compare new method (red)
  with the brute-force calculation using \eq{lnL} without
  approximations (black). Different  line-styles
  represent different number of SNe: $N_{sn} = 1000$ (solid),
  $N_{sn} = 5000$ (dashed) and $N_{sn} = 10000$ (dotted). $P_\alpha$ 
stands for $P_\alpha = (f^2({\bar z})P(k,\bar{z}))_\alpha$. 
The normalization $L_0$ is chosen such that $L/L_0$ peaks at around 1. }
\label{s8-Nsn}
\end{figure}

Results are shown in the \fig{s8-Nsn}, where we plot the
probability distributions for $\sigma_8$ for different number of
SNe. On the vertical line we plotted the relative probability $L$.  As
expected, with increasing number of SNe the distributions become more
alike. If we fitted Gaussian distributions we found that the variances
of the distributions for $N_{sn} = 5000$ differ by roughly $40\%$,
while the variances for $N_{sn} = 10000$ differ only by a few percent
($\sim 4\%$). We stress that the point of this exercise is to see how
many supernovae are required for reaching the Gaussian limit and that
in general one should marginalise over other parameters. We also note
that for any given realization, the maximum likelihood is expected to
be distributed around the fiducial value according to the measurement
error.

\begin{figure}
\centerline{\includegraphics[width=8.0cm]{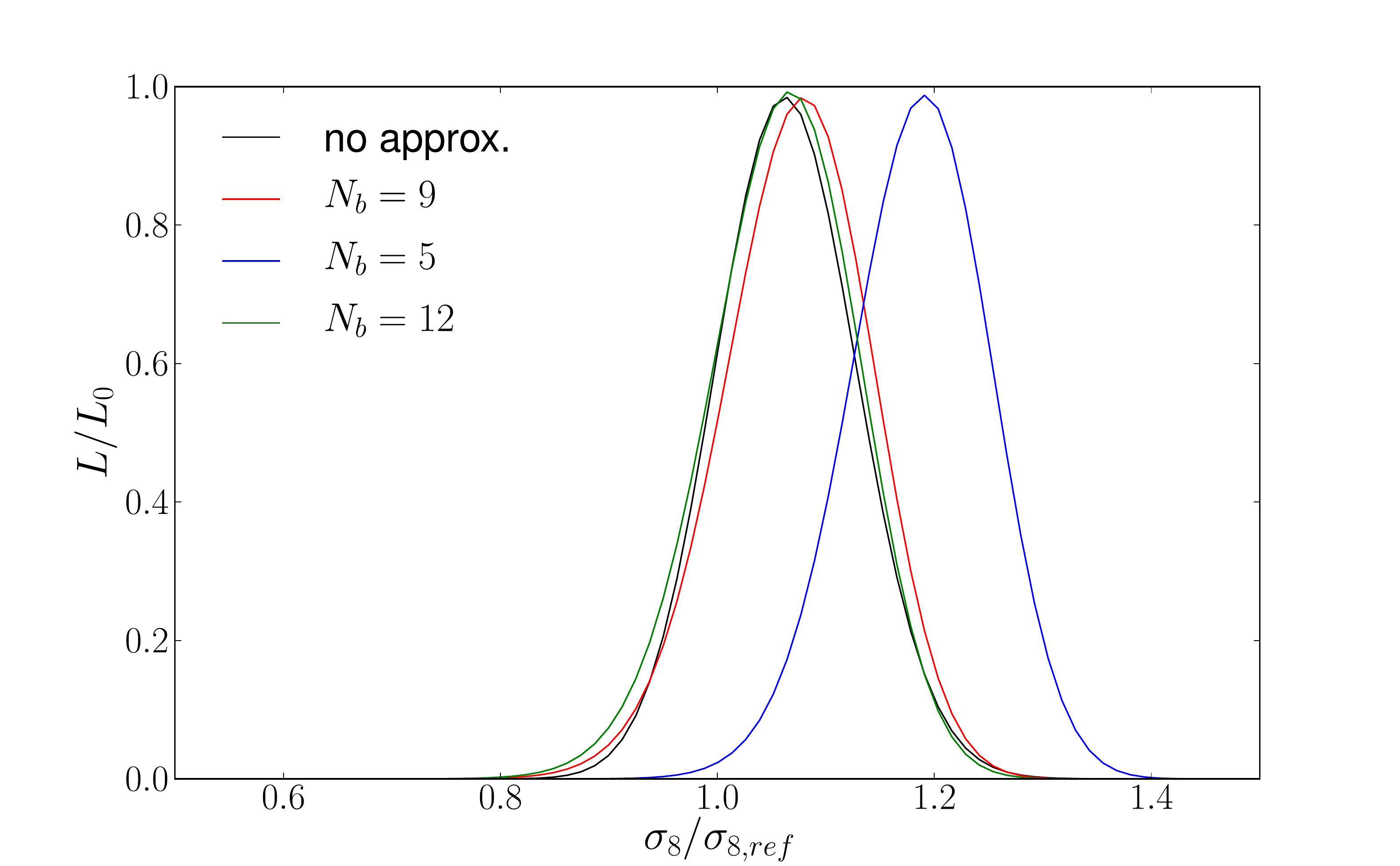}}
\caption{ Shows the probability distribution of cosmological parameter
  $\sigma_8$ for data $P_\alpha$ with respect to the reference power
  spectrum. Different line colours represent different number of bins
  at the constant number of SNe $N_{sn} = 10000$: $N_b = 9$ (red),
  $N_b = 5$ (blue) and $N_b = 12$ (green). Black colour represents the
  brute-force calculation without using any approximations. $P_\alpha$ 
stands for $P_\alpha = (f^2({\bar z})P(k,\bar{z}))_\alpha$. 
The normalization $L_0$ is chosen such that $L/L_0$ peaks at around 1.  }
\label{s8-bins}
\end{figure}

Using the same method we checked how many bins are required. We plot
this in the Figure (\ref{s8-bins}). The distributions vary in shape a little
when
changing number of bins as well as their positions in the k-space. But
as long as we fix the number of bins around $N_b\sim10$ the variances
of the distributions vary for only a few percent.

We now turn to the dependence on the assumption of the fiducial
cosmological model.  To this end we generate synthetic data with
cosmology A ($\Omega_m = 0.24, w = -1$) and reconstruct the matter
power spectrum $P_\alpha$ with the same cosmology. Then we reconstruct
the $P_\alpha$ with different cosmology $B$ ($B\neq A$) and compare
the results with respect to the error on the $P_\alpha$ bins
reconstructed with A.

\begin{figure*}
\centering
\begin{tabular}{cc}
\includegraphics[width=8.0cm]{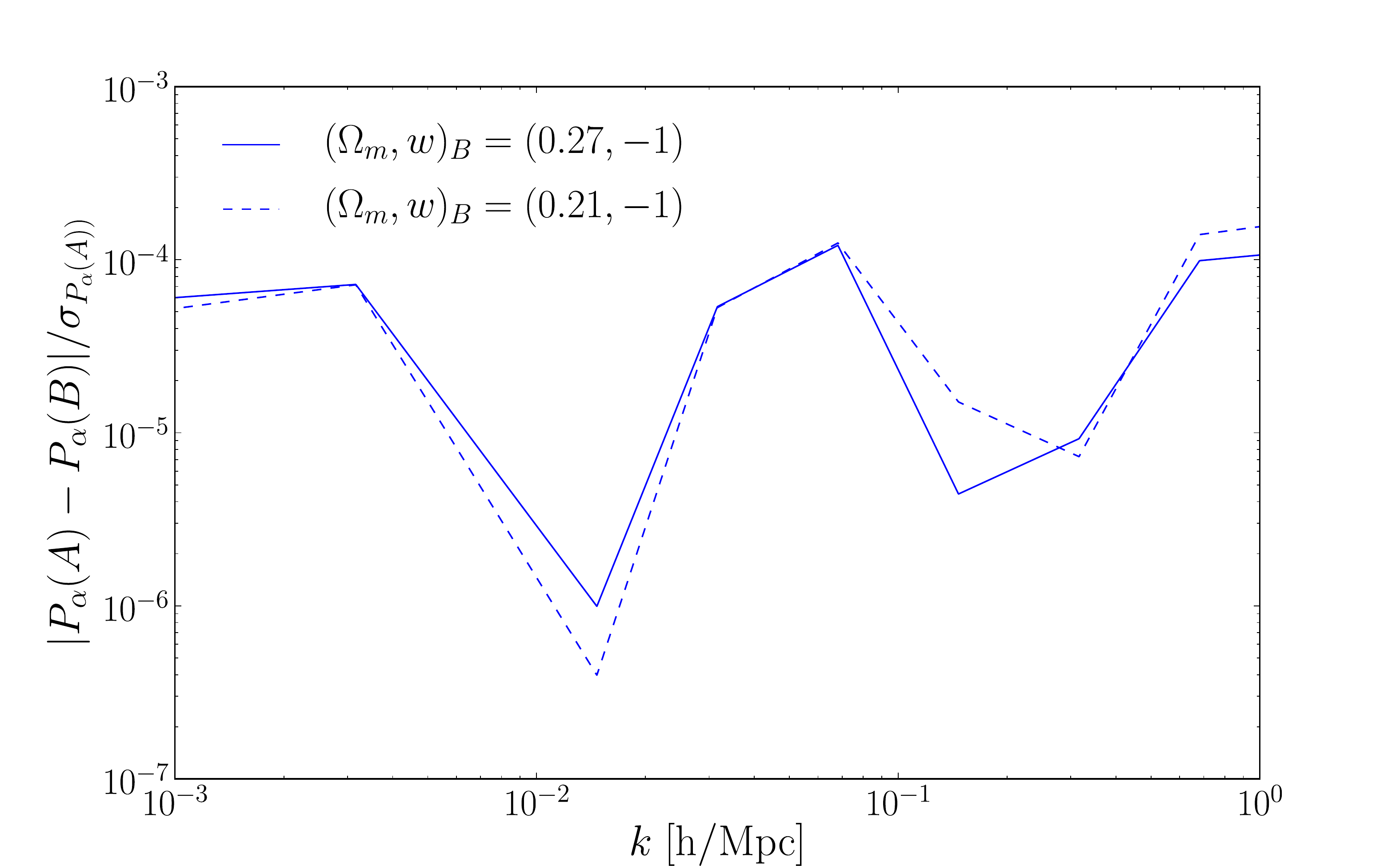} &
\includegraphics[width=8.0cm]{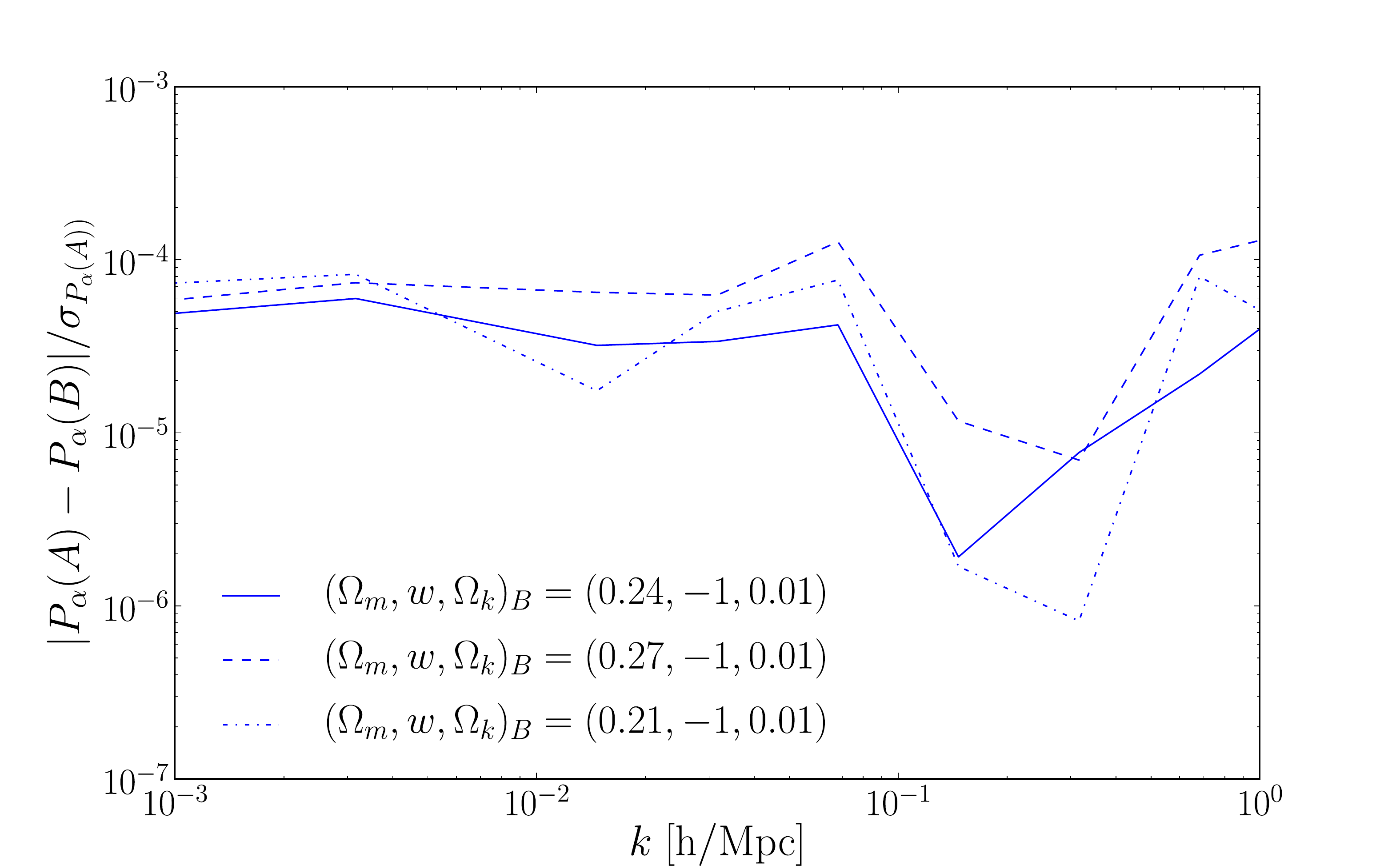}\\
\includegraphics[width=8.0cm]{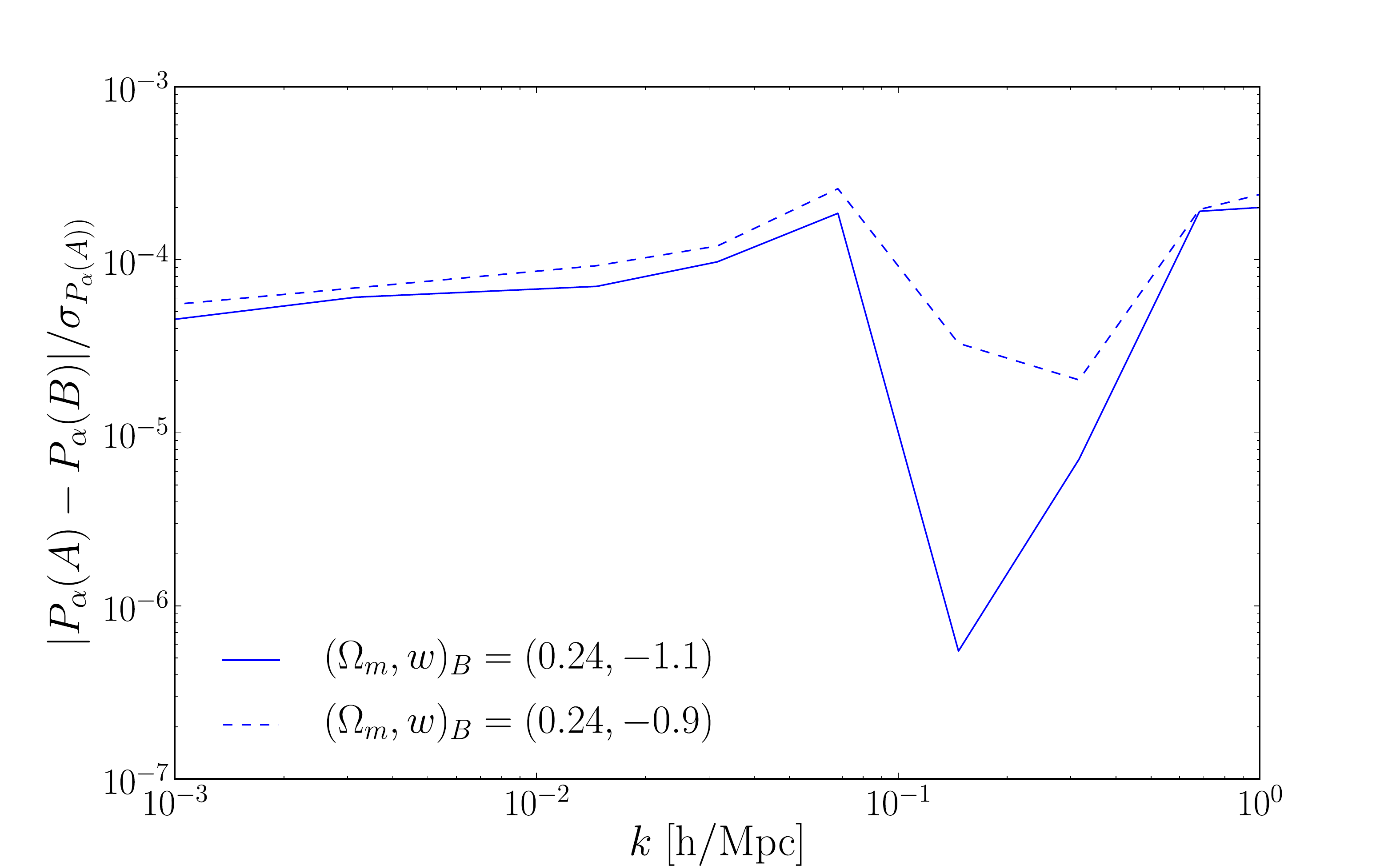} &
\includegraphics[width=8.0cm]{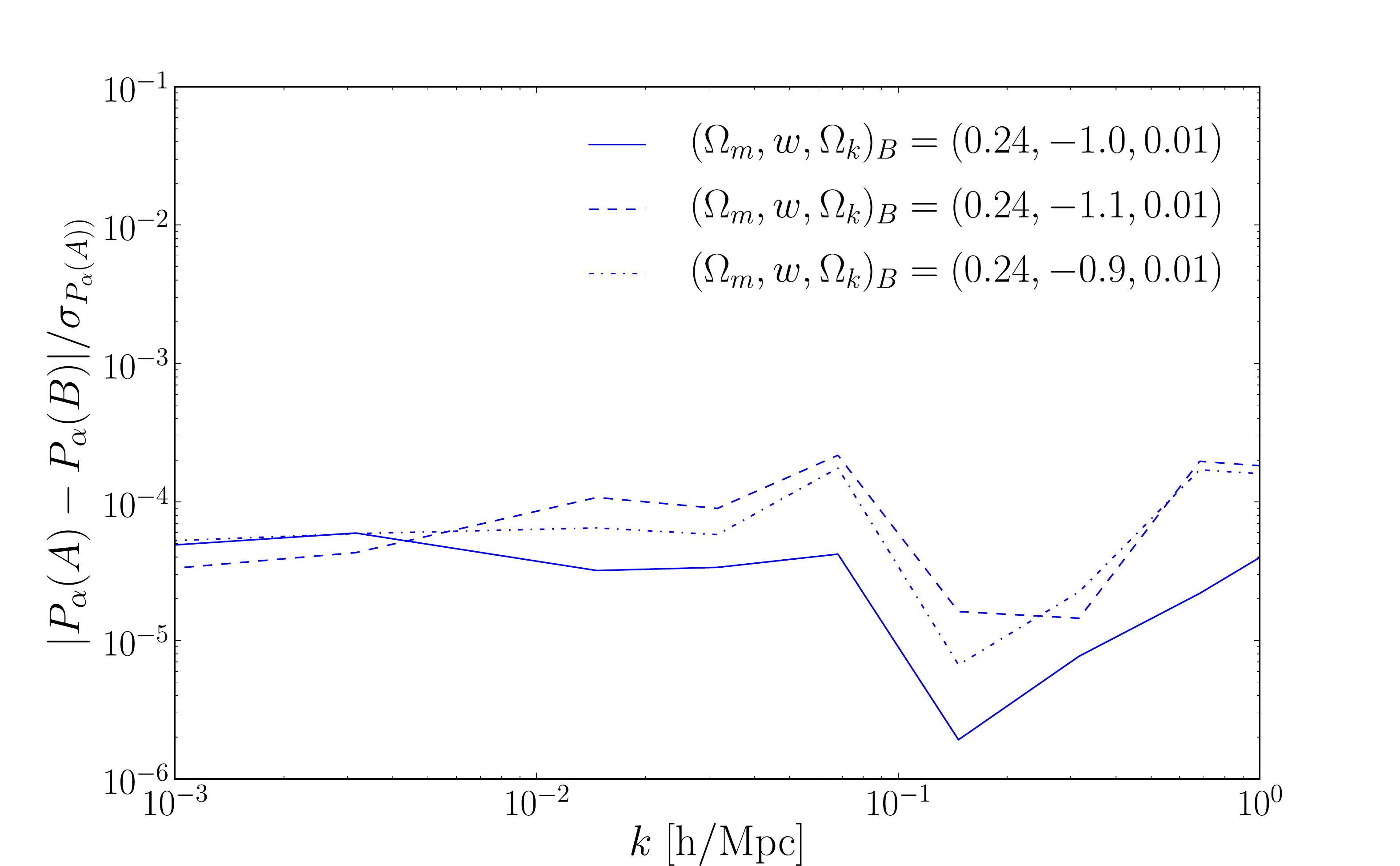}\\
\end{tabular}
\caption{ Shows the difference of the $P_\alpha$ bins at different
  cosmologies ($A$ and $B$) with respect to the error on the result
  reconstructed with $A$. On the upper left plot we change $\Omega_m$
  while keeping $w$ constant ($\Omega_m = 0.27$ - full line, $\Omega_m
  = 0.21$ - dashed; cosmology $A$ had $\Omega_m = 0.24$). On the
  lower left plot  we change $w$ while keeping $\Omega_m$ constant
  ($w=-1.1$ - full line, $w=-0.9$ - dashed; cosmology $A$ had
  $w=-1.0$). The right-hand side plots are the same as left, but we
  have additionally used $\Omega_k=0.01$ in the assumed model.
  }
\label{cosmo-test}
\end{figure*}

Figure \ref{cosmo-test} shows that the results vary with the input
cosmology only in the order of a few percent of the error and are thus
negligible, when we vary parameter $\Omega_m$ and $w$ within their
present error-bars. Which shows that even if we use the slightly wrong
fiducial cosmology in the power spectrum recovery procedure, this does
not affect the result significantly and that was the purpose of this
new method. In this way the most cosmological information is encoded
in the result ($((f\,D)^2 P)_\alpha$).

\begin{figure}
\centerline{\includegraphics[width=8.0cm]{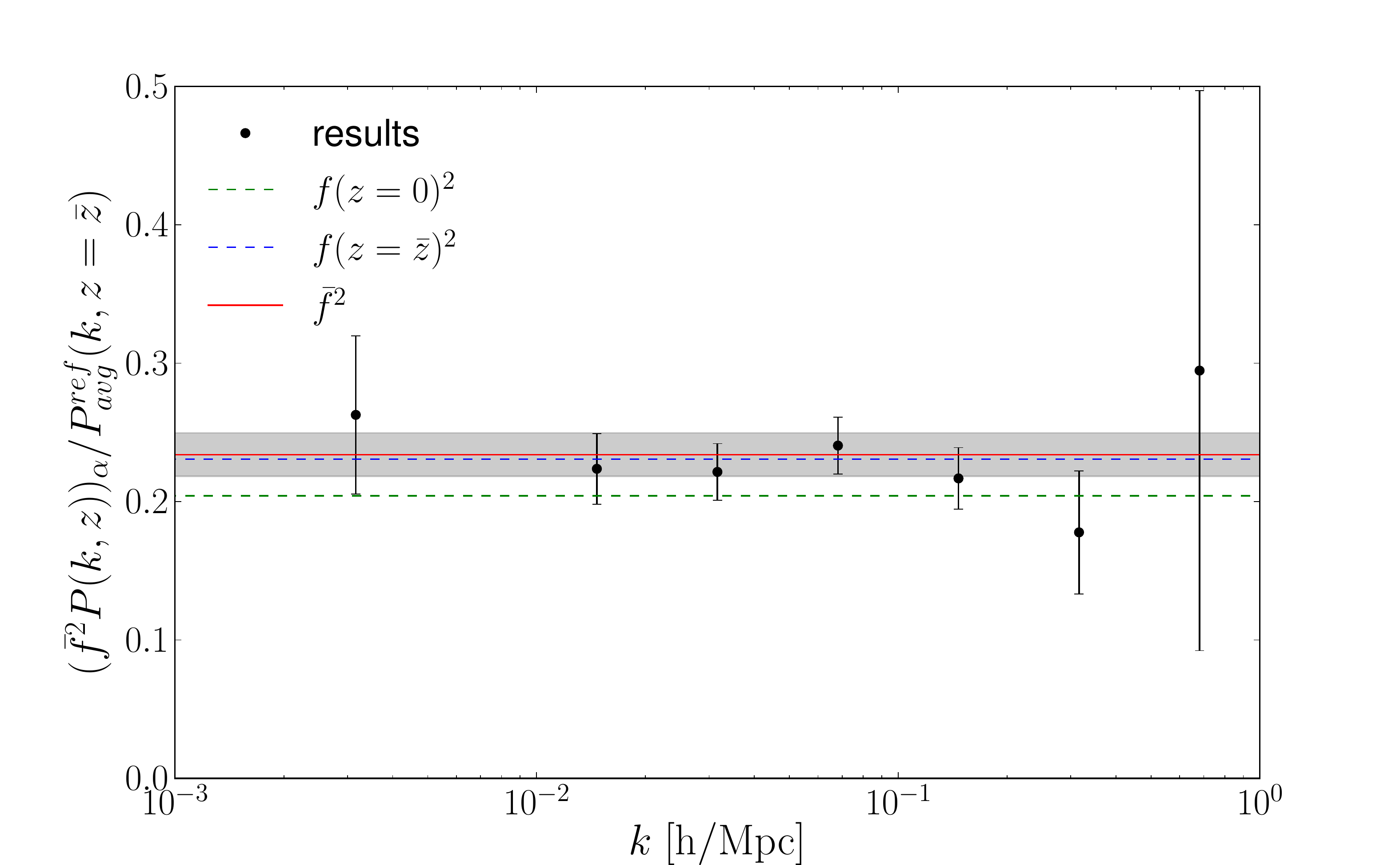}}
\caption{ Shows the results of the new data compression method in the
  form $({\bar f}^2 P(k,z))_\alpha/P^{ref}_{avg}(z=\bar{z})$ on the
  $k_\alpha$ bins. Two dashed constant lines represent values of $f$
  evaluated at two different redshifts ($z=0$ - green and $z = {\bar
    z}$ - blue; where $\bar{z}$ stands for the mean redshift of the
  synthetic SNe data). The mean of the results is shown as a full red
  line with $1\sigma$ errors (gray shaded region). The results are
  consistent with the latter ($f = f(z=\bar{z})$). We used the OQE
  with $N_b = 9$ bins and 10 realizations with $N_{sn} = 10,000$
  SNe.   }
\label{fbar}
\end{figure}

The last cross-check that we perform is to check if result can indeed
be approximated as if all supernovae were at a mean redshift.  The
matter power spectrum and $f$ indeed vary with the redshift. In order
to proceed, we divide the result of our method with average matter power 
spectrum for each bin ($P^{ref}_{avg}(z=\bar{z})$), where $\bar{z}$ is the mean
redshift of the input SNe synthetic data. If the quantity
$P^{ref}_{avg}(z=\bar{z})$ is a good estimator of what we are
measuring we should get a constant line at the value of $f^2$. This is
shown on figure (\ref{fbar}), where we used 9 bins ($N_b = 9$) and 10
realizations with $N_{sn} = 10,000$ SNe.  We see that indeed the right
$f$ to use is the one evaluated at the mean redshift $f = f(z={\bar
  z})$.

\fig{s8-Nsn} shows that future surveys of several thousand SNe Ia will
be able to estimate $\sigma_8$ to the accuracy of $6\%$ using $10,000$ 
SNe Ia.

Finally, we have investigated to what redshift we can push our method,
by creating mock catalogs with increasing maximum redshfit. The
approximations employed in this work clearly start to break down at
z=0.2 and therefore the data need to be split into several redshift
bins if working over $z=0.1$. 

---

\section{Conclusions}\label{conclusion}
\label{sec:conclusions}

To sum up, we presented a new data compression method for cosmological
extraction of data from the low-z SNe. We have shown that these
supernovae effectively measure $P(k,z)f(z)^2$ at the mean redshift of
the supernovae survey and that in the limit of large number of
supernova $N\sim 10,000$, the method is essentially optimal and
unbiased. We have further shown that the correlations of supernovae
velocities are sourced by perturbations on scales with wave-vectors
$\sim10^{-2}$Mpc/$h$ to $\sim0.5$Mpc/$h$ as these scales are the most
constrained. 

Using mock catalogues we have shown that accuracy of $6\%$
in $\sigma_8$ can be obtained using $10,000$ SNe Ia with the data compression
method presented in this paper. We would also like to stress that using velocity
angular power spectrum
should be excercised with caution, since $C_\ell$'s at different redshift
bins are not correlated as was shown in section \ref{sec:proj-veloc-angul}.
Moreover, the data compression method presented 
in this paper does not carry any new information, it just shows that all
information is represented by $f^2(\bar{z})\,P(k,\bar{z})$. 

These basic conclusions are unlikely to change with the real
data. However, at the limit of this number of supernovae, one should
worry about potential systematics and biases. Two of those are the most
important. First, the actual perturbations sourcing supernovae are not
in the completely linear regime any more and second, supernovae don't
trace velocity at random positions in the Universe, but instead trace
it at position of rare density peaks in the primordial field, where
galaxies eventually form. Corrections due to these effects can only be
assessed by running $N$-body simulations. This exceeds to scope of
this paper.

Recent SNe data contain of the order of a couple hundred low-redshift
supernovae and so this method remains inapplicable. Unfortunately, the
$10,000$ low-redshift SNe will not be available for the foreseeable
future. The LSST survey predicts $\sim 100,000$ SNe per year with
redshift $z<1.2$. Of this $\sim 1400$ SNe per year with
$z<0.1$. However, there is no planned spectroscopic followup to these
measurements. The LSST photometric redshift errors are predicted to be
around $0.02 (1+z)$.  We examined if information on peculiar
velocities could be extracted with such large redshift errors by
approximating them with an extra velocity scatter of $6000$km/s, but
the results are not encouraging. Photometric errors of course have an
additional issue of being highly non-Gaussian.

If one could, however, measure 10,000 supernova Ia at low redshift, it
might results in some interesting science. The auto-power spectrum of
the population of galaxies in which supernova reside would measure the
value of $P(k)b^2$, where $b$ is the galaxy bias, while the
redshift-space distortions of the same sample would measure $\beta =
f/b$. Combined with peculiar velocities measurements discussed here,
one would get an over-constrained system and if analysis was done on
the same volume, the sample variance would cancel out.  The quantity
\be S_G = \frac{1}{\beta^2}\frac{P_g}{P_{sn}}, \label{sg} \ee would
therefore have a unity value in Einstein gravity and deviation from
unity would indicate either new physics or systematics.  Note that in
using several tracers over the same field, the sample variance would
cancel and thus provide a much more stringent result than one might
naively guess. Similar method using weak gravitational lensing was recently used
in (\cite{2010Natur.464..256R}).

---


\section*{Acknowledgements}

VI acknowledges support of the Berkeley Center of Cosmological
Physics, where parts of this work were completed during summer working
visit. AS is supported in part by the U.S. Department of Energy under
Contract No. DE-AC02-98CH10886.

-

\vspace{0.5cm}
\section*{Appendix}

\renewcommand{\theequation}{A\arabic{equation}}
\setcounter{equation}{0}

When computing the angular power spectrum  we get
\begin{align}
\langle a_{\ell m} a^*_{\ell'm'} \rangle &= \bar{D}'(z_1) \bar{D}'(z_2) \int
\frac{k^2 dk}{(2\pi)^3} P(k) \int d\Omega_k \frac{1}{k^4} \cdot \notag \\ 
& \cdot \int d\Omega(\hbx_1) Y^*_{\ell m}(\hbx_1) e^{i \bk \cdot \bx_1} \bk
\cdot \hbx_1 \cdot \notag \\
& \cdot \int d\Omega(\hbx_2) Y_{\ell' m'}(\hbx_2) e^{-i \bk \cdot \bx_2} \bk
\cdot \hbx_2,
\label{app_kotni_spekter}
\end{align}
where $a_{\ell m}$ stand for the expansion coefficients, $\bar{D}(z)$
is normalized linear growth factor and derivatives are with respect to
conformal time ($a(t)d\eta = cdt$). In equation
(\ref{app_kotni_spekter}) there are two integrals of the form \be
I_{\ell m} = \int d\Omega(\hbx) Y^*_{\ell m}(\hbx) e^{i \bk \cdot \bx}
\bk \cdot \hbx,
\label{app_1}
\ee where we have dropped the SN label in the subscript (1 or 2). We
notice that the integrals in \eq{app_kotni_spekter} are complex
conjugates, so we only need to calculate one. If we
choose the coordinate system such that $\bk = k \left(
  \sin{\theta_k}\cos{\phi_k},\sin{\theta_k}\sin{\phi_k},\cos{\theta_k}
\right)$ and $\bx = x \left(
  \sin{\theta}\cos{\phi},\sin{\theta}\sin{\phi},\cos{\theta} \right)$
we can rewrite the above integral into \be I_{\ell m} = \int d\Omega
Y^*_{\ell m}(\theta,\phi) e^{i kx \cos{\gamma}} k\cos{\gamma},
\label{app_2}
\ee
where $k = |\bk|$, $x = |\bx|$ and $\gamma$ the angle between $\bk$ in $\bx$.
Also the following identity is true
\be
\frac{k}{ix}\frac{\partial}{\partial k} e^{ikx\cos{\gamma}} =
e^{ikx\cos{\gamma}} k \cos{\gamma}.
\ee
Now we are left with the integral of the exponent function over the spherical
harmonics. With the use of the following mathematical identities
\cite{arfken+weber}
\be
e^{ikr\cos{\gamma}} = \sum_{n=0}^{\infty} a_n j_n(kx) P_n(\cos{\gamma}),
\ee
where $a_n = i^n (2n+1)$, $j_n$ are spherical Bessel functions and $P_n$
Legendre polynomials given by
\be
P_n(\cos{\gamma}) = \frac{4\pi}{2n +
1}\sum_{p=-n}^{+n}Y_{np}(\theta,\phi)Y_{np}^*(\theta_k,\phi_k),
\ee
where $Y_{np}$ are spherical harmonics, we can rewrite \eq{app_2} into
\begin{align}
I_{\ell m} =& \frac{k}{ix} \frac{\partial}{\partial k} \sum_{n=0}^{\infty} a_n
\frac{4\pi}{2n+1} j_n(kx) \cdot \notag \\ 
& \cdot \sum_{p=-n}^{+n} Y_{np}*(\theta_k,\phi_k) \int d\Omega Y_{\ell
m}^*(\theta,\phi)Y_{np}(\theta,\phi),
\label{app_3}
\end{align}
where the only integral left is the orthonormal relation of spherical harmonics
\be
\int d\Omega Y_{\ell m}^*(\theta,\phi)Y_{np}(\theta,\phi) = \delta_{\ell
n}\delta_{mp},
\ee
where $\delta$'s on the right-hand side are Kronecker delta's. Using the above
identity we can evaluate both sums in the \eq{app_3} and write the result as
\begin{align}
I_{\ell m} &= \frac{k}{ix} \frac{\partial}{\partial k} a_{\ell}
\frac{4\pi}{2\ell + 1} j_{\ell}(kx)Y^*_{\ell m}(\theta_k,\phi_k) \notag \\
&= 4\pi \,k\,i^{\ell-1} \frac{\partial j_{\ell}(kx)}{x\partial k} Y^*_{\ell
m}(\theta_k,\phi_k).
\end{align}
Comparing the result with the original equation (\ref{app_kotni_spekter}), we
have computed $I_{\ell m}(kx_1)$. With complex conjugation and variable
substitution of this result we can write the second integral as $I_{\ell'
m'}^*(kx_2)$, where $x_{1,2}=|\bx_{1,2}|$. Rewriting \eq{app_kotni_spekter} and
inserting the integral values we get
\begin{align}
\langle a_{\ell m} a^*_{\ell'm'} \rangle =& \bar{D}'(z_1) \bar{D}'(z_2) \int
\frac{k^2 dk}{(2\pi)^3} P(k) \cdot \notag \\
& \cdot \int d\Omega_k \frac{1}{k^4} I_{\ell m}(kx_1) I_{\ell' m'}(kx_2) \notag
\\
=& \bar{D}'(z_1) \bar{D}'(z_2) \int \frac{2dk}{\pi} P(k) \cdot \notag \\
& \cdot \frac{\partial j_{\ell}(kx_1)}{x_1\partial k} \frac{\partial
j_{\ell'}(kx_2)}{x_2\partial k} i^{\ell - \ell'} \cdot \notag \\
& \cdot \int d\Omega_k Y^*_{\ell m}(\theta_k,\phi_k) Y_{\ell'
m'}(\theta_k,\phi_k) \notag \\
=& \bar{D}'(z_1) \bar{D}'(z_2) \int \frac{2dk}{\pi} P(k) \cdot \notag \\
& \cdot \frac{\partial j_{\ell}(kx_1)}{x_1\partial k} \frac{\partial
j_{\ell'}(kx_2)}{x_2\partial k} i^{\ell - \ell'} \delta_{\ell \ell'} \delta_{m
m'},
\end{align}
which is exactly what we were looking for. The term on the right-hand side of
the equation, that stands in front of the $\delta_{\ell \ell'} \delta_{m m'}$ is
by definition equal to $C_\ell$. Implying the Kronecker delta $\delta_{\ell
\ell'}$ we can write
\begin{align}
C_\ell(z_1,z_2) =& \bar{D}'(z_1) \bar{D}'(z_2) \frac{2}{\pi} \cdot \notag \\
& \cdot \int_0^\infty dk P(k) \left( \frac{\partial j_\ell(kx_1)}{x_1\partial k}
\right) \left( \frac{\partial j_\ell(kx_2)}{x_2\partial k} \right).
\end{align}

\bibliography{peculiarSN}

\begin{thebibliography}{10}%
\makeatletter
\providecommand \@ifxundefined [1]{%
 \ifx #1\undefined \expandafter \@firstoftwo
 \else \expandafter \@secondoftwo
\fi
}%
\providecommand \@ifnum [1]{%
 \ifnum #1\expandafter \@firstoftwo
 \else \expandafter \@secondoftwo
\fi
}%
\providecommand \enquote [1]{``#1''}%
\providecommand \bibnamefont  [1]{#1}%
\providecommand \bibfnamefont [1]{#1}%
\providecommand \citenamefont [1]{#1}%
\providecommand\href[0]{\@sanitize\@href}%
\providecommand\@href[1]{\endgroup\@@startlink{#1}\endgroup\@@href}%
\providecommand\@@href[1]{#1\@@endlink}%
\providecommand \@sanitize [0]{\begingroup\catcode`\&12\catcode`\#12\relax}%
\@ifxundefined \pdfoutput {\@firstoftwo}{%
 \@ifnum{\z@=\pdfoutput}{\@firstoftwo}{\@secondoftwo}%
}{%
 \providecommand\@@startlink[1]{\leavevmode\special{html:<a href="#1">}}%
 \providecommand\@@endlink[0]{\special{html:</a>}}%
}{%
 \providecommand\@@startlink[1]{%
  \leavevmode
  \pdfstartlink
   attr{/Border[0 0 1 ]/H/I/C[0 1 1]}%
   user{/Subtype/Link/A<</Type/Action/S/URI/URI(#1)>>}%
  \relax
 }%
 \providecommand\@@endlink[0]{\pdfendlink}%
}%
\providecommand \url  [0]{\begingroup\@sanitize \@url }%
\providecommand \@url [1]{\endgroup\@href {#1}{\urlprefix}}%
\providecommand \urlprefix [0]{URL }%
\providecommand \Eprint[0]{\href }%
\@ifxundefined \urlstyle {%
  \providecommand \doi [1]{doi:\discretionary{}{}{}#1}%
}{%
  \providecommand \doi [0]{doi:\discretionary{}{}{}\begingroup
  \urlstyle{rm}\Url }%
}%
\providecommand \doibase [0]{http://dx.doi.org/}%
\providecommand \Doi[1]{\href{\doibase#1}}%
\providecommand \bibAnnote [3]{%
  \BibitemShut{#1}%
  \begin{quotation}\noindent
    \textsc{Key:}\ #2\\\textsc{Annotation:}\ #3%
  \end{quotation}%
}%
\providecommand \bibAnnoteFile [2]{%
  \IfFileExists{#2}{\bibAnnote {#1} {#2} {\input{#2}}}{}%
}%
\providecommand \typeout [0]{\immediate \write \m@ne }%
\providecommand \selectlanguage [0]{\@gobble}%
\providecommand \bibinfo [0]{\@secondoftwo}%
\providecommand \bibfield [0]{\@secondoftwo}%
\providecommand \translation [1]{[#1]}%
\providecommand \BibitemOpen[0]{}%
\providecommand \bibitemStop [0]{}%
\providecommand \bibitemNoStop [0]{.\EOS\space}%
\providecommand \EOS [0]{\spacefactor3000\relax}%
\providecommand \BibitemShut [1]{\csname bibitem#1\endcsname}%
\bibitem{1998AJ....116.1009R}%
  \BibitemOpen
  \bibfield{author}{%
  \bibinfo {author} {\bibfnamefont{A.~G.}\ \bibnamefont{{Riess}}}, \bibinfo
  {author} {\bibfnamefont{A.~V.}\ \bibnamefont{{Filippenko}}}, \bibinfo
  {author} {\bibfnamefont{P.}~\bibnamefont{{Challis}}}, \bibinfo {author}
  {\bibfnamefont{A.}~\bibnamefont{{Clocchiatti}}}, \bibinfo {author}
  {\bibfnamefont{A.}~\bibnamefont{{Diercks}}}, \bibinfo {author}
  {\bibfnamefont{P.~M.}\ \bibnamefont{{Garnavich}}}, \bibinfo {author}
  {\bibfnamefont{R.~L.}\ \bibnamefont{{Gilliland}}}, \bibinfo {author}
  {\bibfnamefont{C.~J.}\ \bibnamefont{{Hogan}}}, \bibinfo {author}
  {\bibfnamefont{S.}~\bibnamefont{{Jha}}}, \bibinfo {author}
  {\bibfnamefont{R.~P.}\ \bibnamefont{{Kirshner}}}, \bibinfo {author}
  {\bibfnamefont{B.}~\bibnamefont{{Leibundgut}}}, \bibinfo {author}
  {\bibfnamefont{M.~M.}\ \bibnamefont{{Phillips}}}, \bibinfo {author}
  {\bibfnamefont{D.}~\bibnamefont{{Reiss}}}, \bibinfo {author}
  {\bibfnamefont{B.~P.}\ \bibnamefont{{Schmidt}}}, \bibinfo {author}
  {\bibfnamefont{R.~A.}\ \bibnamefont{{Schommer}}}, \bibinfo {author}
  {\bibfnamefont{R.~C.}\ \bibnamefont{{Smith}}}, \bibinfo {author}
  {\bibfnamefont{J.}~\bibnamefont{{Spyromilio}}}, \bibinfo {author}
  {\bibfnamefont{C.}~\bibnamefont{{Stubbs}}}, \bibinfo {author}
  {\bibfnamefont{N.~B.}\ \bibnamefont{{Suntzeff}}},\ and\ \bibinfo {author}
  {\bibfnamefont{J.}~\bibnamefont{{Tonry}}},\ }%
  \bibfield{journal}{%
  \Doi{10.1086/300499}{\bibinfo {journal} {\aj}}\ }%
  \textbf{\bibinfo {volume} {116}},\ \bibinfo {pages} {1009} (\bibinfo {month}
  {Sep.}\ \bibinfo {year} {1998}),\
  \Eprint{http://arxiv.org/abs/arXiv:astro-ph/9805201}{arXiv:astro-ph/9805201}%
  \bibAnnoteFile{NoStop}{1998AJ....116.1009R}%
\bibitem{1999ApJ...517..565P}%
  \BibitemOpen
  \bibfield{author}{%
  \bibinfo {author} {\bibfnamefont{S.}~\bibnamefont{{Perlmutter}}}, \bibinfo
  {author} {\bibfnamefont{G.}~\bibnamefont{{Aldering}}}, \bibinfo {author}
  {\bibfnamefont{G.}~\bibnamefont{{Goldhaber}}}, \bibinfo {author}
  {\bibfnamefont{R.~A.}\ \bibnamefont{{Knop}}}, \bibinfo {author}
  {\bibfnamefont{P.}~\bibnamefont{{Nugent}}}, \bibinfo {author}
  {\bibfnamefont{P.~G.}\ \bibnamefont{{Castro}}}, \bibinfo {author}
  {\bibfnamefont{S.}~\bibnamefont{{Deustua}}}, \bibinfo {author}
  {\bibfnamefont{S.}~\bibnamefont{{Fabbro}}}, \bibinfo {author}
  {\bibfnamefont{A.}~\bibnamefont{{Goobar}}}, \bibinfo {author}
  {\bibfnamefont{D.~E.}\ \bibnamefont{{Groom}}}, \bibinfo {author}
  {\bibfnamefont{I.~M.}\ \bibnamefont{{Hook}}}, \bibinfo {author}
  {\bibfnamefont{A.~G.}\ \bibnamefont{{Kim}}}, \bibinfo {author}
  {\bibfnamefont{M.~Y.}\ \bibnamefont{{Kim}}}, \bibinfo {author}
  {\bibfnamefont{J.~C.}\ \bibnamefont{{Lee}}}, \bibinfo {author}
  {\bibfnamefont{N.~J.}\ \bibnamefont{{Nunes}}}, \bibinfo {author}
  {\bibfnamefont{R.}~\bibnamefont{{Pain}}}, \bibinfo {author}
  {\bibfnamefont{C.~R.}\ \bibnamefont{{Pennypacker}}}, \bibinfo {author}
  {\bibfnamefont{R.}~\bibnamefont{{Quimby}}}, \bibinfo {author}
  {\bibfnamefont{C.}~\bibnamefont{{Lidman}}}, \bibinfo {author}
  {\bibfnamefont{R.~S.}\ \bibnamefont{{Ellis}}}, \bibinfo {author}
  {\bibfnamefont{M.}~\bibnamefont{{Irwin}}}, \bibinfo {author}
  {\bibfnamefont{R.~G.}\ \bibnamefont{{McMahon}}}, \bibinfo {author}
  {\bibfnamefont{P.}~\bibnamefont{{Ruiz-Lapuente}}}, \bibinfo {author}
  {\bibfnamefont{N.}~\bibnamefont{{Walton}}}, \bibinfo {author}
  {\bibfnamefont{B.}~\bibnamefont{{Schaefer}}}, \bibinfo {author}
  {\bibfnamefont{B.~J.}\ \bibnamefont{{Boyle}}}, \bibinfo {author}
  {\bibfnamefont{A.~V.}\ \bibnamefont{{Filippenko}}}, \bibinfo {author}
  {\bibfnamefont{T.}~\bibnamefont{{Matheson}}}, \bibinfo {author}
  {\bibfnamefont{A.~S.}\ \bibnamefont{{Fruchter}}}, \bibinfo {author}
  {\bibfnamefont{N.}~\bibnamefont{{Panagia}}}, \bibinfo {author}
  {\bibfnamefont{H.~J.~M.}\ \bibnamefont{{Newberg}}}, \bibinfo {author}
  {\bibfnamefont{W.~J.}\ \bibnamefont{{Couch}}},\ and\ \bibinfo {author}
  {\bibnamefont{{The Supernova Cosmology Project}}},\ }%
  \bibfield{journal}{%
  \Doi{10.1086/307221}{\bibinfo {journal} {\apj}}\ }%
  \textbf{\bibinfo {volume} {517}},\ \bibinfo {pages} {565} (\bibinfo {month}
  {Jun.}\ \bibinfo {year} {1999}),\
  \Eprint{http://arxiv.org/abs/arXiv:astro-ph/9812133}{arXiv:astro-ph/9812133}%
  \bibAnnoteFile{NoStop}{1999ApJ...517..565P}%
\bibitem{2006A&A...447...31A}%
  \BibitemOpen
  \bibfield{author}{%
  \bibinfo {author} {\bibfnamefont{P.}~\bibnamefont{{Astier}}}, \bibinfo
  {author} {\bibfnamefont{J.}~\bibnamefont{{Guy}}}, \bibinfo {author}
  {\bibfnamefont{N.}~\bibnamefont{{Regnault}}}, \bibinfo {author}
  {\bibfnamefont{R.}~\bibnamefont{{Pain}}}, \bibinfo {author}
  {\bibfnamefont{E.}~\bibnamefont{{Aubourg}}}, \bibinfo {author}
  {\bibfnamefont{D.}~\bibnamefont{{Balam}}}, \bibinfo {author}
  {\bibfnamefont{S.}~\bibnamefont{{Basa}}}, \bibinfo {author}
  {\bibfnamefont{R.~G.}\ \bibnamefont{{Carlberg}}}, \bibinfo {author}
  {\bibfnamefont{S.}~\bibnamefont{{Fabbro}}}, \bibinfo {author}
  {\bibfnamefont{D.}~\bibnamefont{{Fouchez}}}, \bibinfo {author}
  {\bibfnamefont{I.~M.}\ \bibnamefont{{Hook}}}, \bibinfo {author}
  {\bibfnamefont{D.~A.}\ \bibnamefont{{Howell}}}, \bibinfo {author}
  {\bibfnamefont{H.}~\bibnamefont{{Lafoux}}}, \bibinfo {author}
  {\bibfnamefont{J.~D.}\ \bibnamefont{{Neill}}}, \bibinfo {author}
  {\bibfnamefont{N.}~\bibnamefont{{Palanque-Delabrouille}}}, \bibinfo {author}
  {\bibfnamefont{K.}~\bibnamefont{{Perrett}}}, \bibinfo {author}
  {\bibfnamefont{C.~J.}\ \bibnamefont{{Pritchet}}}, \bibinfo {author}
  {\bibfnamefont{J.}~\bibnamefont{{Rich}}}, \bibinfo {author}
  {\bibfnamefont{M.}~\bibnamefont{{Sullivan}}}, \bibinfo {author}
  {\bibfnamefont{R.}~\bibnamefont{{Taillet}}}, \bibinfo {author}
  {\bibfnamefont{G.}~\bibnamefont{{Aldering}}}, \bibinfo {author}
  {\bibfnamefont{P.}~\bibnamefont{{Antilogus}}}, \bibinfo {author}
  {\bibfnamefont{V.}~\bibnamefont{{Arsenijevic}}}, \bibinfo {author}
  {\bibfnamefont{C.}~\bibnamefont{{Balland}}}, \bibinfo {author}
  {\bibfnamefont{S.}~\bibnamefont{{Baumont}}}, \bibinfo {author}
  {\bibfnamefont{J.}~\bibnamefont{{Bronder}}}, \bibinfo {author}
  {\bibfnamefont{H.}~\bibnamefont{{Courtois}}}, \bibinfo {author}
  {\bibfnamefont{R.~S.}\ \bibnamefont{{Ellis}}}, \bibinfo {author}
  {\bibfnamefont{M.}~\bibnamefont{{Filiol}}}, \bibinfo {author}
  {\bibfnamefont{A.~C.}\ \bibnamefont{{Gon{\c c}alves}}}, \bibinfo {author}
  {\bibfnamefont{A.}~\bibnamefont{{Goobar}}}, \bibinfo {author}
  {\bibfnamefont{D.}~\bibnamefont{{Guide}}}, \bibinfo {author}
  {\bibfnamefont{D.}~\bibnamefont{{Hardin}}}, \bibinfo {author}
  {\bibfnamefont{V.}~\bibnamefont{{Lusset}}}, \bibinfo {author}
  {\bibfnamefont{C.}~\bibnamefont{{Lidman}}}, \bibinfo {author}
  {\bibfnamefont{R.}~\bibnamefont{{McMahon}}}, \bibinfo {author}
  {\bibfnamefont{M.}~\bibnamefont{{Mouchet}}}, \bibinfo {author}
  {\bibfnamefont{A.}~\bibnamefont{{Mourao}}}, \bibinfo {author}
  {\bibfnamefont{S.}~\bibnamefont{{Perlmutter}}}, \bibinfo {author}
  {\bibfnamefont{P.}~\bibnamefont{{Ripoche}}}, \bibinfo {author}
  {\bibfnamefont{C.}~\bibnamefont{{Tao}}},\ and\ \bibinfo {author}
  {\bibfnamefont{N.}~\bibnamefont{{Walton}}},\ }%
  \bibfield{journal}{%
  \Doi{10.1051/0004-6361:20054185}{\bibinfo {journal} {\aap}}\ }%
  \textbf{\bibinfo {volume} {447}},\ \bibinfo {pages} {31} (\bibinfo {month}
  {Feb.}\ \bibinfo {year} {2006}),\
  \Eprint{http://arxiv.org/abs/arXiv:astro-ph/0510447}{arXiv:astro-ph/0510447}%
  \bibAnnoteFile{NoStop}{2006A&A...447...31A}%
\bibitem{2007ApJ...659...98R}%
  \BibitemOpen
  \bibfield{author}{%
  \bibinfo {author} {\bibfnamefont{A.~G.}\ \bibnamefont{{Riess}}}, \bibinfo
  {author} {\bibfnamefont{L.}~\bibnamefont{{Strolger}}}, \bibinfo {author}
  {\bibfnamefont{S.}~\bibnamefont{{Casertano}}}, \bibinfo {author}
  {\bibfnamefont{H.~C.}\ \bibnamefont{{Ferguson}}}, \bibinfo {author}
  {\bibfnamefont{B.}~\bibnamefont{{Mobasher}}}, \bibinfo {author}
  {\bibfnamefont{B.}~\bibnamefont{{Gold}}}, \bibinfo {author}
  {\bibfnamefont{P.~J.}\ \bibnamefont{{Challis}}}, \bibinfo {author}
  {\bibfnamefont{A.~V.}\ \bibnamefont{{Filippenko}}}, \bibinfo {author}
  {\bibfnamefont{S.}~\bibnamefont{{Jha}}}, \bibinfo {author}
  {\bibfnamefont{W.}~\bibnamefont{{Li}}}, \bibinfo {author}
  {\bibfnamefont{J.}~\bibnamefont{{Tonry}}}, \bibinfo {author}
  {\bibfnamefont{R.}~\bibnamefont{{Foley}}}, \bibinfo {author}
  {\bibfnamefont{R.~P.}\ \bibnamefont{{Kirshner}}}, \bibinfo {author}
  {\bibfnamefont{M.}~\bibnamefont{{Dickinson}}}, \bibinfo {author}
  {\bibfnamefont{E.}~\bibnamefont{{MacDonald}}}, \bibinfo {author}
  {\bibfnamefont{D.}~\bibnamefont{{Eisenstein}}}, \bibinfo {author}
  {\bibfnamefont{M.}~\bibnamefont{{Livio}}}, \bibinfo {author}
  {\bibfnamefont{J.}~\bibnamefont{{Younger}}}, \bibinfo {author}
  {\bibfnamefont{C.}~\bibnamefont{{Xu}}}, \bibinfo {author}
  {\bibfnamefont{T.}~\bibnamefont{{Dahl{\'e}n}}},\ and\ \bibinfo {author}
  {\bibfnamefont{D.}~\bibnamefont{{Stern}}},\ }%
  \bibfield{journal}{%
  \Doi{10.1086/510378}{\bibinfo {journal} {\apj}}\ }%
  \textbf{\bibinfo {volume} {659}},\ \bibinfo {pages} {98} (\bibinfo {month}
  {Apr.}\ \bibinfo {year} {2007}),\
  \Eprint{http://arxiv.org/abs/arXiv:astro-ph/0611572}{arXiv:astro-ph/0611572}%
  \bibAnnoteFile{NoStop}{2007ApJ...659...98R}%
\bibitem{2007ApJ...666..694W}%
  \BibitemOpen
  \bibfield{author}{%
  \bibinfo {author} {\bibfnamefont{W.~M.}\ \bibnamefont{{Wood-Vasey}}},
  \bibinfo {author} {\bibfnamefont{G.}~\bibnamefont{{Miknaitis}}}, \bibinfo
  {author} {\bibfnamefont{C.~W.}\ \bibnamefont{{Stubbs}}}, \bibinfo {author}
  {\bibfnamefont{S.}~\bibnamefont{{Jha}}}, \bibinfo {author}
  {\bibfnamefont{A.~G.}\ \bibnamefont{{Riess}}}, \bibinfo {author}
  {\bibfnamefont{P.~M.}\ \bibnamefont{{Garnavich}}}, \bibinfo {author}
  {\bibfnamefont{R.~P.}\ \bibnamefont{{Kirshner}}}, \bibinfo {author}
  {\bibfnamefont{C.}~\bibnamefont{{Aguilera}}}, \bibinfo {author}
  {\bibfnamefont{A.~C.}\ \bibnamefont{{Becker}}}, \bibinfo {author}
  {\bibfnamefont{J.~W.}\ \bibnamefont{{Blackman}}}, \bibinfo {author}
  {\bibfnamefont{S.}~\bibnamefont{{Blondin}}}, \bibinfo {author}
  {\bibfnamefont{P.}~\bibnamefont{{Challis}}}, \bibinfo {author}
  {\bibfnamefont{A.}~\bibnamefont{{Clocchiatti}}}, \bibinfo {author}
  {\bibfnamefont{A.}~\bibnamefont{{Conley}}}, \bibinfo {author}
  {\bibfnamefont{R.}~\bibnamefont{{Covarrubias}}}, \bibinfo {author}
  {\bibfnamefont{T.~M.}\ \bibnamefont{{Davis}}}, \bibinfo {author}
  {\bibfnamefont{A.~V.}\ \bibnamefont{{Filippenko}}}, \bibinfo {author}
  {\bibfnamefont{R.~J.}\ \bibnamefont{{Foley}}}, \bibinfo {author}
  {\bibfnamefont{A.}~\bibnamefont{{Garg}}}, \bibinfo {author}
  {\bibfnamefont{M.}~\bibnamefont{{Hicken}}}, \bibinfo {author}
  {\bibfnamefont{K.}~\bibnamefont{{Krisciunas}}}, \bibinfo {author}
  {\bibfnamefont{B.}~\bibnamefont{{Leibundgut}}}, \bibinfo {author}
  {\bibfnamefont{W.}~\bibnamefont{{Li}}}, \bibinfo {author}
  {\bibfnamefont{T.}~\bibnamefont{{Matheson}}}, \bibinfo {author}
  {\bibfnamefont{A.}~\bibnamefont{{Miceli}}}, \bibinfo {author}
  {\bibfnamefont{G.}~\bibnamefont{{Narayan}}}, \bibinfo {author}
  {\bibfnamefont{G.}~\bibnamefont{{Pignata}}}, \bibinfo {author}
  {\bibfnamefont{J.~L.}\ \bibnamefont{{Prieto}}}, \bibinfo {author}
  {\bibfnamefont{A.}~\bibnamefont{{Rest}}}, \bibinfo {author}
  {\bibfnamefont{M.~E.}\ \bibnamefont{{Salvo}}}, \bibinfo {author}
  {\bibfnamefont{B.~P.}\ \bibnamefont{{Schmidt}}}, \bibinfo {author}
  {\bibfnamefont{R.~C.}\ \bibnamefont{{Smith}}}, \bibinfo {author}
  {\bibfnamefont{J.}~\bibnamefont{{Sollerman}}}, \bibinfo {author}
  {\bibfnamefont{J.}~\bibnamefont{{Spyromilio}}}, \bibinfo {author}
  {\bibfnamefont{J.~L.}\ \bibnamefont{{Tonry}}}, \bibinfo {author}
  {\bibfnamefont{N.~B.}\ \bibnamefont{{Suntzeff}}},\ and\ \bibinfo {author}
  {\bibfnamefont{A.}~\bibnamefont{{Zenteno}}},\ }%
  \bibfield{journal}{%
  \Doi{10.1086/518642}{\bibinfo {journal} {\apj}}\ }%
  \textbf{\bibinfo {volume} {666}},\ \bibinfo {pages} {694} (\bibinfo {month}
  {Sep.}\ \bibinfo {year} {2007}),\
  \Eprint{http://arxiv.org/abs/arXiv:astro-ph/0701041}{arXiv:astro-ph/0701041}%
  \bibAnnoteFile{NoStop}{2007ApJ...666..694W}%
\bibitem{2002SPIE.4836.....T}%
  \BibitemOpen
  \bibinfo {editor} {\bibfnamefont{J.~A.}\ \bibnamefont{{Tyson}}}\ and\
  \bibinfo {editor} {\bibfnamefont{S.}~\bibnamefont{{Wolff}}},\ eds.,\
  \emph{\bibinfo {title} {Society of Photo-Optical Instrumentation Engineers
  (SPIE) Conference Series}},\ \bibinfo {series} {Presented at the Society of
  Photo-Optical Instrumentation Engineers (SPIE) Conference}, Vol.\ \bibinfo
  {volume} {4836}\ (\bibinfo {year} {2002})%
  \bibAnnoteFile{NoStop}{2002SPIE.4836.....T}%
\bibitem{2005AAS...20718005D}%
  \BibitemOpen
  \bibfield{author}{%
  \bibinfo {author} {\bibfnamefont{B.}~\bibnamefont{{Dilday}}}, \bibinfo
  {author} {\bibfnamefont{J.}~\bibnamefont{{Barentine}}}, \bibinfo {author}
  {\bibfnamefont{B.}~\bibnamefont{{Bassett}}}, \bibinfo {author}
  {\bibfnamefont{A.}~\bibnamefont{{Becker}}}, \bibinfo {author}
  {\bibfnamefont{R.}~\bibnamefont{{Bendar}}}, \bibinfo {author}
  {\bibfnamefont{M.}~\bibnamefont{{Bremer}}}, \bibinfo {author}
  {\bibfnamefont{H.}~\bibnamefont{{Brewington}}}, \bibinfo {author}
  {\bibfnamefont{F.}~\bibnamefont{{DeJongh}}}, \bibinfo {author}
  {\bibfnamefont{J.}~\bibnamefont{{Dembicky}}}, \bibinfo {author}
  {\bibfnamefont{D.~L.}\ \bibnamefont{{DePoy}}}, \bibinfo {author}
  {\bibfnamefont{M.}~\bibnamefont{{Doi}}}, \bibinfo {author}
  {\bibfnamefont{A.}~\bibnamefont{{Edge}}}, \bibinfo {author}
  {\bibfnamefont{E.}~\bibnamefont{{Elson}}}, \bibinfo {author}
  {\bibfnamefont{J.}~\bibnamefont{{Frieman}}}, \bibinfo {author}
  {\bibfnamefont{P.}~\bibnamefont{{Garnavich}}}, \bibinfo {author}
  {\bibfnamefont{A.}~\bibnamefont{{Goobar}}}, \bibinfo {author}
  {\bibfnamefont{T.}~\bibnamefont{{Gueth}}}, \bibinfo {author}
  {\bibfnamefont{M.}~\bibnamefont{{Harvanek}}}, \bibinfo {author}
  {\bibfnamefont{J.}~\bibnamefont{{Holtzman}}}, \bibinfo {author}
  {\bibfnamefont{U.}~\bibnamefont{{Hopp}}}, \bibinfo {author}
  {\bibfnamefont{W.}~\bibnamefont{{Kollatschny}}}, \bibinfo {author}
  {\bibfnamefont{J.}~\bibnamefont{{Krzesinski}}}, \bibinfo {author}
  {\bibfnamefont{D.}~\bibnamefont{{Lamenti}}}, \bibinfo {author}
  {\bibfnamefont{H.}~\bibnamefont{{Lampeitl}}}, \bibinfo {author}
  {\bibfnamefont{R.}~\bibnamefont{{Kessler}}}, \bibinfo {author}
  {\bibfnamefont{B.}~\bibnamefont{{Ketzeback}}}, \bibinfo {author}
  {\bibfnamefont{K.}~\bibnamefont{{Konishi}}}, \bibinfo {author}
  {\bibfnamefont{D.}~\bibnamefont{{Long}}}, \bibinfo {author}
  {\bibfnamefont{J.}~\bibnamefont{{Marriner}}}, \bibinfo {author}
  {\bibfnamefont{J.~L.}\ \bibnamefont{{Marshall}}}, \bibinfo {author}
  {\bibfnamefont{R.}~\bibnamefont{{McMillan}}}, \bibinfo {author}
  {\bibfnamefont{J.}~\bibnamefont{{Mendez}}}, \bibinfo {author}
  {\bibfnamefont{G.}~\bibnamefont{{Miknaitis}}}, \bibinfo {author}
  {\bibfnamefont{R.}~\bibnamefont{{Nichol}}}, \bibinfo {author}
  {\bibfnamefont{K.}~\bibnamefont{{Pan}}}, \bibinfo {author}
  {\bibfnamefont{J.~L.}\ \bibnamefont{{Prieto}}}, \bibinfo {author}
  {\bibfnamefont{M.}~\bibnamefont{{Richmond}}}, \bibinfo {author}
  {\bibfnamefont{A.}~\bibnamefont{{Riess}}}, \bibinfo {author}
  {\bibfnamefont{R.}~\bibnamefont{{Romani}}}, \bibinfo {author}
  {\bibfnamefont{K.}~\bibnamefont{{Romer}}}, \bibinfo {author}
  {\bibfnamefont{P.}~\bibnamefont{{Ruiz-Lapuente}}}, \bibinfo {author}
  {\bibfnamefont{M.}~\bibnamefont{{Sako}}}, \bibinfo {author}
  {\bibfnamefont{D.}~\bibnamefont{{Schneider}}}, \bibinfo {author}
  {\bibfnamefont{M.}~\bibnamefont{{Smith}}}, \bibinfo {author}
  {\bibfnamefont{S.}~\bibnamefont{{Snedden}}}, \bibinfo {author}
  {\bibfnamefont{M.}~\bibnamefont{{Subbarao}}}, \bibinfo {author}
  {\bibfnamefont{N.}~\bibnamefont{{Takanashi}}}, \bibinfo {author}
  {\bibfnamefont{K.}~\bibnamefont{{van der Heyden}}}, \bibinfo {author}
  {\bibfnamefont{C.}~\bibnamefont{{Wheeler}}},\ and\ \bibinfo {author}
  {\bibfnamefont{N.}~\bibnamefont{{Yasuda}}},\ }%
  in\ \emph{\bibinfo {booktitle} {Bulletin of the American Astronomical
  Society}},\ \bibinfo {series} {Bulletin of the American Astronomical
  Society}, Vol.~\bibinfo {volume} {37}\ (\bibinfo {year} {2005})\ pp.\
  \bibinfo {pages} {1459--+}%
  \bibAnnoteFile{NoStop}{2005AAS...20718005D}%
\bibitem{2006PASP..118....2H}%
  \BibitemOpen
  \bibfield{author}{%
  \bibinfo {author} {\bibfnamefont{M.}~\bibnamefont{{Hamuy}}}, \bibinfo
  {author} {\bibfnamefont{G.}~\bibnamefont{{Folatelli}}}, \bibinfo {author}
  {\bibfnamefont{N.~I.}\ \bibnamefont{{Morrell}}}, \bibinfo {author}
  {\bibfnamefont{M.~M.}\ \bibnamefont{{Phillips}}}, \bibinfo {author}
  {\bibfnamefont{N.~B.}\ \bibnamefont{{Suntzeff}}}, \bibinfo {author}
  {\bibfnamefont{S.~E.}\ \bibnamefont{{Persson}}}, \bibinfo {author}
  {\bibfnamefont{M.}~\bibnamefont{{Roth}}}, \bibinfo {author}
  {\bibfnamefont{S.}~\bibnamefont{{Gonzalez}}}, \bibinfo {author}
  {\bibfnamefont{W.}~\bibnamefont{{Krzeminski}}}, \bibinfo {author}
  {\bibfnamefont{C.}~\bibnamefont{{Contreras}}}, \bibinfo {author}
  {\bibfnamefont{W.~L.}\ \bibnamefont{{Freedman}}}, \bibinfo {author}
  {\bibfnamefont{D.~C.}\ \bibnamefont{{Murphy}}}, \bibinfo {author}
  {\bibfnamefont{B.~F.}\ \bibnamefont{{Madore}}}, \bibinfo {author}
  {\bibfnamefont{P.}~\bibnamefont{{Wyatt}}}, \bibinfo {author}
  {\bibfnamefont{J.}~\bibnamefont{{Maza}}}, \bibinfo {author}
  {\bibfnamefont{A.~V.}\ \bibnamefont{{Filippenko}}}, \bibinfo {author}
  {\bibfnamefont{W.}~\bibnamefont{{Li}}},\ and\ \bibinfo {author}
  {\bibfnamefont{P.~A.}\ \bibnamefont{{Pinto}}},\ }%
  \bibfield{journal}{%
  \Doi{10.1086/500228}{\bibinfo {journal} {\pasp}}\ }%
  \textbf{\bibinfo {volume} {118}},\ \bibinfo {pages} {2} (\bibinfo {month}
  {Jan.}\ \bibinfo {year} {2006}),\
  \Eprint{http://arxiv.org/abs/arXiv:astro-ph/0512039}{arXiv:astro-ph/0512039}%
  \bibAnnoteFile{NoStop}{2006PASP..118....2H}%
\bibitem{2005NewAR..49..346A}%
  \BibitemOpen
  \bibfield{author}{%
  \bibinfo {author} {\bibfnamefont{G.}~\bibnamefont{{Aldering}}},\ }%
  \bibfield{journal}{%
  \Doi{10.1016/j.newar.2005.08.002}{\bibinfo {journal} {\nar}}\ }%
  \textbf{\bibinfo {volume} {49}},\ \bibinfo {pages} {346} (\bibinfo {month}
  {Nov.}\ \bibinfo {year} {2005}),\
  \Eprint{http://arxiv.org/abs/arXiv:astro-ph/0507426}{arXiv:astro-ph/0507426}%
  \bibAnnoteFile{NoStop}{2005NewAR..49..346A}%
\bibitem{1995ApJ...445L..91R}%
  \BibitemOpen
  \bibfield{author}{%
  \bibinfo {author} {\bibfnamefont{A.~G.}\ \bibnamefont{{Riess}}}, \bibinfo
  {author} {\bibfnamefont{W.~H.}\ \bibnamefont{{Press}}},\ and\ \bibinfo
  {author} {\bibfnamefont{R.~P.}\ \bibnamefont{{Kirshner}}},\ }%
  \bibfield{journal}{%
  \Doi{10.1086/187897}{\bibinfo {journal} {\apjl}}\ }%
  \textbf{\bibinfo {volume} {445}},\ \bibinfo {pages} {L91} (\bibinfo {month}
  {Jun.}\ \bibinfo {year} {1995}),\
  \Eprint{http://arxiv.org/abs/arXiv:astro-ph/9412017}{arXiv:astro-ph/9412017}%
  \bibAnnoteFile{NoStop}{1995ApJ...445L..91R}%
\bibitem{1997ApJ...488L...1R}%
  \BibitemOpen
  \bibfield{author}{%
  \bibinfo {author} {\bibfnamefont{A.~G.}\ \bibnamefont{{Riess}}}, \bibinfo
  {author} {\bibfnamefont{M.}~\bibnamefont{{Davis}}}, \bibinfo {author}
  {\bibfnamefont{J.}~\bibnamefont{{Baker}}},\ and\ \bibinfo {author}
  {\bibfnamefont{R.~P.}\ \bibnamefont{{Kirshner}}},\ }%
  \bibfield{journal}{%
  \Doi{10.1086/310917}{\bibinfo {journal} {\apjl}}\ }%
  \textbf{\bibinfo {volume} {488}},\ \bibinfo {pages} {L1+} (\bibinfo {month}
  {Oct.}\ \bibinfo {year} {1997}),\
  \Eprint{http://arxiv.org/abs/arXiv:astro-ph/9707261}{arXiv:astro-ph/9707261}%
  \bibAnnoteFile{NoStop}{1997ApJ...488L...1R}%
\bibitem{1998ApJ...503..483Z}%
  \BibitemOpen
  \bibfield{author}{%
  \bibinfo {author} {\bibfnamefont{I.}~\bibnamefont{{Zehavi}}}, \bibinfo
  {author} {\bibfnamefont{A.~G.}\ \bibnamefont{{Riess}}}, \bibinfo {author}
  {\bibfnamefont{R.~P.}\ \bibnamefont{{Kirshner}}},\ and\ \bibinfo {author}
  {\bibfnamefont{A.}~\bibnamefont{{Dekel}}},\ }%
  \bibfield{journal}{%
  \Doi{10.1086/306015}{\bibinfo {journal} {\apj}}\ }%
  \textbf{\bibinfo {volume} {503}},\ \bibinfo {pages} {483} (\bibinfo {month}
  {Aug.}\ \bibinfo {year} {1998}),\
  \Eprint{http://arxiv.org/abs/arXiv:astro-ph/9802252}{arXiv:astro-ph/9802252}%
  \bibAnnoteFile{NoStop}{1998ApJ...503..483Z}%
\bibitem{2000AAS...196.6207B}%
  \BibitemOpen
  \bibfield{author}{%
  \bibinfo {author} {\bibfnamefont{A.}~\bibnamefont{{Bonacic}}}, \bibinfo
  {author} {\bibfnamefont{R.~A.}\ \bibnamefont{{Schommer}}}, \bibinfo {author}
  {\bibfnamefont{N.~B.}\ \bibnamefont{{Suntzeff}}},\ and\ \bibinfo {author}
  {\bibfnamefont{M.~M.}\ \bibnamefont{{Phillips}}},\ }%
  in\ \emph{\bibinfo {booktitle} {Bulletin of the American Astronomical
  Society}},\ \bibinfo {series} {Bulletin of the American Astronomical
  Society}, Vol.~\bibinfo {volume} {32}\ (\bibinfo {year} {2000})\ pp.\
  \bibinfo {pages} {1285--+}%
  \bibAnnoteFile{NoStop}{2000AAS...196.6207B}%
\bibitem{2004MNRAS.355.1378R}%
  \BibitemOpen
  \bibfield{author}{%
  \bibinfo {author} {\bibfnamefont{D.~J.}\ \bibnamefont{{Radburn-Smith}}},
  \bibinfo {author} {\bibfnamefont{J.~R.}\ \bibnamefont{{Lucey}}},\ and\
  \bibinfo {author} {\bibfnamefont{M.~J.}\ \bibnamefont{{Hudson}}},\ }%
  \bibfield{journal}{%
  \Doi{10.1111/j.1365-2966.2004.08420.x}{\bibinfo {journal} {\mnras}}\ }%
  \textbf{\bibinfo {volume} {355}},\ \bibinfo {pages} {1378} (\bibinfo {month}
  {Dec.}\ \bibinfo {year} {2004}),\
  \Eprint{http://arxiv.org/abs/arXiv:astro-ph/0409551}{arXiv:astro-ph/0409551}%
  \bibAnnoteFile{NoStop}{2004MNRAS.355.1378R}%
\bibitem{2006PhRvL..96s1302B}%
  \BibitemOpen
  \bibfield{author}{%
  \bibinfo {author} {\bibfnamefont{C.}~\bibnamefont{{Bonvin}}}, \bibinfo
  {author} {\bibfnamefont{R.}~\bibnamefont{{Durrer}}},\ and\ \bibinfo {author}
  {\bibfnamefont{M.}~\bibnamefont{{Kunz}}},\ }%
  \bibfield{journal}{%
  \Doi{10.1103/PhysRevLett.96.191302}{\bibinfo {journal} {Physical Review
  Letters}}\ }%
  \textbf{\bibinfo {volume} {96}},\ \bibinfo {pages} {191302} (\bibinfo {month}
  {May}\ \bibinfo {year} {2006}),\
  \Eprint{http://arxiv.org/abs/arXiv:astro-ph/0603240}{arXiv:astro-ph/0603240}%
  \bibAnnoteFile{NoStop}{2006PhRvL..96s1302B}%
\bibitem{2007ApJ...661..650H}%
  \BibitemOpen
  \bibfield{author}{%
  \bibinfo {author} {\bibfnamefont{T.}~\bibnamefont{{Haugb{\o}lle}}}, \bibinfo
  {author} {\bibfnamefont{S.}~\bibnamefont{{Hannestad}}}, \bibinfo {author}
  {\bibfnamefont{B.}~\bibnamefont{{Thomsen}}}, \bibinfo {author}
  {\bibfnamefont{J.}~\bibnamefont{{Fynbo}}}, \bibinfo {author}
  {\bibfnamefont{J.}~\bibnamefont{{Sollerman}}},\ and\ \bibinfo {author}
  {\bibfnamefont{S.}~\bibnamefont{{Jha}}},\ }%
  \bibfield{journal}{%
  \Doi{10.1086/513600}{\bibinfo {journal} {\apj}}\ }%
  \textbf{\bibinfo {volume} {661}},\ \bibinfo {pages} {650} (\bibinfo {month}
  {Jun.}\ \bibinfo {year} {2007}),\
  \Eprint{http://arxiv.org/abs/arXiv:astro-ph/0612137}{arXiv:astro-ph/0612137}%
  \bibAnnoteFile{NoStop}{2007ApJ...661..650H}%
\bibitem{2007ApJ...659..122J}%
  \BibitemOpen
  \bibfield{author}{%
  \bibinfo {author} {\bibfnamefont{S.}~\bibnamefont{{Jha}}}, \bibinfo {author}
  {\bibfnamefont{A.~G.}\ \bibnamefont{{Riess}}},\ and\ \bibinfo {author}
  {\bibfnamefont{R.~P.}\ \bibnamefont{{Kirshner}}},\ }%
  \bibfield{journal}{%
  \Doi{10.1086/512054}{\bibinfo {journal} {\apj}}\ }%
  \textbf{\bibinfo {volume} {659}},\ \bibinfo {pages} {122} (\bibinfo {month}
  {Apr.}\ \bibinfo {year} {2007}),\
  \Eprint{http://arxiv.org/abs/arXiv:astro-ph/0612666}{arXiv:astro-ph/0612666}%
  \bibAnnoteFile{NoStop}{2007ApJ...659..122J}%
\bibitem{2007MNRAS.379..343W}%
  \BibitemOpen
  \bibfield{author}{%
  \bibinfo {author} {\bibfnamefont{R.}~\bibnamefont{{Watkins}}}\ and\ \bibinfo
  {author} {\bibfnamefont{H.~A.}\ \bibnamefont{{Feldman}}},\ }%
  \bibfield{journal}{%
  \Doi{10.1111/j.1365-2966.2007.11970.x}{\bibinfo {journal} {\mnras}}\ }%
  \textbf{\bibinfo {volume} {379}},\ \bibinfo {pages} {343} (\bibinfo {month}
  {Jul.}\ \bibinfo {year} {2007}),\
  \Eprint{http://arxiv.org/abs/arXiv:astro-ph/0702751}{arXiv:astro-ph/0702751}%
  \bibAnnoteFile{NoStop}{2007MNRAS.379..343W}%
\bibitem{2007ApJ...664L..13C}%
  \BibitemOpen
  \bibfield{author}{%
  \bibinfo {author} {\bibfnamefont{A.}~\bibnamefont{{Conley}}}, \bibinfo
  {author} {\bibfnamefont{R.~G.}\ \bibnamefont{{Carlberg}}}, \bibinfo {author}
  {\bibfnamefont{J.}~\bibnamefont{{Guy}}}, \bibinfo {author}
  {\bibfnamefont{D.~A.}\ \bibnamefont{{Howell}}}, \bibinfo {author}
  {\bibfnamefont{S.}~\bibnamefont{{Jha}}}, \bibinfo {author}
  {\bibfnamefont{A.~G.}\ \bibnamefont{{Riess}}},\ and\ \bibinfo {author}
  {\bibfnamefont{M.}~\bibnamefont{{Sullivan}}},\ }%
  \bibfield{journal}{%
  \Doi{10.1086/520625}{\bibinfo {journal} {\apjl}}\ }%
  \textbf{\bibinfo {volume} {664}},\ \bibinfo {pages} {L13} (\bibinfo {month}
  {Jul.}\ \bibinfo {year} {2007}),\
  \Eprint{http://arxiv.org/abs/0705.0367}{arXiv:0705.0367}%
  \bibAnnoteFile{NoStop}{2007ApJ...664L..13C}%
\bibitem{2007arXiv0705.0368W}%
  \BibitemOpen
  \bibfield{author}{%
  \bibinfo {author} {\bibfnamefont{L.}~\bibnamefont{{Wang}}},\ }%
  \bibfield{journal}{%
  \bibinfo {journal} {ArXiv e-prints}}%
   (\bibinfo {month} {May}\ \bibinfo {year} {2007}),\
  \Eprint{http://arxiv.org/abs/0705.0368}{arXiv:0705.0368}%
  \bibAnnoteFile{NoStop}{2007arXiv0705.0368W}%
\bibitem{2007ApJ...661L.123N}%
  \BibitemOpen
  \bibfield{author}{%
  \bibinfo {author} {\bibfnamefont{J.~D.}\ \bibnamefont{{Neill}}}, \bibinfo
  {author} {\bibfnamefont{M.~J.}\ \bibnamefont{{Hudson}}},\ and\ \bibinfo
  {author} {\bibfnamefont{A.}~\bibnamefont{{Conley}}},\ }%
  \bibfield{journal}{%
  \Doi{10.1086/518808}{\bibinfo {journal} {\apjl}}\ }%
  \textbf{\bibinfo {volume} {661}},\ \bibinfo {pages} {L123} (\bibinfo {month}
  {Jun.}\ \bibinfo {year} {2007}),\
  \Eprint{http://arxiv.org/abs/0704.1654}{arXiv:0704.1654}%
  \bibAnnoteFile{NoStop}{2007ApJ...661L.123N}%
\bibitem{2008JCAP...02..022H}%
  \BibitemOpen
  \bibfield{author}{%
  \bibinfo {author} {\bibfnamefont{S.}~\bibnamefont{{Hannestad}}}, \bibinfo
  {author} {\bibfnamefont{T.}~\bibnamefont{{Haugb{\o}lle}}},\ and\ \bibinfo
  {author} {\bibfnamefont{B.}~\bibnamefont{{Thomsen}}},\ }%
  \bibfield{journal}{%
  \Doi{10.1088/1475-7516/2008/02/022}{\bibinfo {journal} {\jcap}}\ }%
  \textbf{\bibinfo {volume} {2}},\ \bibinfo {pages} {22} (\bibinfo {month}
  {Feb.}\ \bibinfo {year} {2008}),\
  \Eprint{http://arxiv.org/abs/0705.0979}{arXiv:0705.0979}%
  \bibAnnoteFile{NoStop}{2008JCAP...02..022H}%
\bibitem{1988ApJ...332L...7G}%
  \BibitemOpen
  \bibfield{author}{%
  \bibinfo {author} {\bibfnamefont{K.}~\bibnamefont{{Gorski}}},\ }%
  \bibfield{journal}{%
  \Doi{10.1086/185255}{\bibinfo {journal} {\apjl}}\ }%
  \textbf{\bibinfo {volume} {332}},\ \bibinfo {pages} {L7} (\bibinfo {month}
  {Sep.}\ \bibinfo {year} {1988})%
  \bibAnnoteFile{NoStop}{1988ApJ...332L...7G}%
\bibitem{2006PhRvD..73l3526H}%
  \BibitemOpen
  \bibfield{author}{%
  \bibinfo {author} {\bibfnamefont{L.}~\bibnamefont{{Hui}}}\ and\ \bibinfo
  {author} {\bibfnamefont{P.~B.}\ \bibnamefont{{Greene}}},\ }%
  \bibfield{journal}{%
  \Doi{10.1103/PhysRevD.73.123526}{\bibinfo {journal} {\prd}}\ }%
  \textbf{\bibinfo {volume} {73}},\ \bibinfo {pages} {123526} (\bibinfo {month}
  {Jun.}\ \bibinfo {year} {2006}),\
  \Eprint{http://arxiv.org/abs/arXiv:astro-ph/0512159}{arXiv:astro-ph/0512159}%
  \bibAnnoteFile{NoStop}{2006PhRvD..73l3526H}%
\bibitem{2007PhRvL..99h1301G}%
  \BibitemOpen
  \bibfield{author}{%
  \bibinfo {author} {\bibfnamefont{C.}~\bibnamefont{{Gordon}}}, \bibinfo
  {author} {\bibfnamefont{K.}~\bibnamefont{{Land}}},\ and\ \bibinfo {author}
  {\bibfnamefont{A.}~\bibnamefont{{Slosar}}},\ }%
  \bibfield{journal}{%
  \Doi{10.1103/PhysRevLett.99.081301}{\bibinfo {journal} {Physical Review
  Letters}}\ }%
  \textbf{\bibinfo {volume} {99}},\ \bibinfo {pages} {081301} (\bibinfo {month}
  {Aug.}\ \bibinfo {year} {2007}),\
  \Eprint{http://arxiv.org/abs/0705.1718}{arXiv:0705.1718}%
  \bibAnnoteFile{NoStop}{2007PhRvL..99h1301G}%
\bibitem{2006PhRvD..73b3523B}%
  \BibitemOpen
  \bibfield{author}{%
  \bibinfo {author} {\bibfnamefont{C.}~\bibnamefont{{Bonvin}}}, \bibinfo
  {author} {\bibfnamefont{R.}~\bibnamefont{{Durrer}}},\ and\ \bibinfo {author}
  {\bibfnamefont{M.~A.}\ \bibnamefont{{Gasparini}}},\ }%
  \bibfield{journal}{%
  \Doi{10.1103/PhysRevD.73.023523}{\bibinfo {journal} {Phys. Rev. D}}\ }%
  \textbf{\bibinfo {volume} {73}},\ \bibinfo {pages} {023523} (\bibinfo {month}
  {Jan.}\ \bibinfo {year} {2006}),\
  \Eprint{http://arxiv.org/abs/arXiv:astro-ph/0511183}{arXiv:astro-ph/0511183}%
  \bibAnnoteFile{NoStop}{2006PhRvD..73b3523B}%
\bibitem{Dodelson}%
  \BibitemOpen
  \bibfield{author}{%
  \bibinfo {author} {\bibfnamefont{S.}~\bibnamefont{Dodelson}},\ }%
  \emph{\bibinfo {title} {Modern Cosmology}}\ (\bibinfo {publisher} {Academic
  Press},\ \bibinfo {year} {2003})\ ISBN \bibinfo {isbn}
  {ISBN-13:987-0-12-219141-1}%
  \bibAnnoteFile{NoStop}{Dodelson}%
\bibitem{Note1}%
  \BibitemOpen
  \bibinfo {note} {Matter power spectrum was computed with CAMB using the
  following cosmological parameters ($\Omega _m = 0.24,\Omega _b = 0.04,h=0.7,w
  = -1, n_s = 1, \sigma _8 = 0.789347$)}%
  \bibAnnoteFile{NoStop}{Note1}%
\bibitem{1998PhRvD..57.2117B}%
  \BibitemOpen
  \bibfield{author}{%
  \bibinfo {author} {\bibfnamefont{J.~R.}\ \bibnamefont{{Bond}}}, \bibinfo
  {author} {\bibfnamefont{A.~H.}\ \bibnamefont{{Jaffe}}},\ and\ \bibinfo
  {author} {\bibfnamefont{L.}~\bibnamefont{{Knox}}},\ }%
  \bibfield{journal}{%
  \Doi{10.1103/PhysRevD.57.2117}{\bibinfo {journal} {Phys. Rev. D}}\ }%
  \textbf{\bibinfo {volume} {57}},\ \bibinfo {pages} {2117} (\bibinfo {month}
  {Feb.}\ \bibinfo {year} {1998}),\
  \Eprint{http://arxiv.org/abs/arXiv:astro-ph/9708203}{arXiv:astro-ph/9708203}%
  \bibAnnoteFile{NoStop}{1998PhRvD..57.2117B}%
\bibitem{2010Natur.464..256R}%
  \BibitemOpen
  \bibfield{author}{%
  \bibinfo {author} {\bibfnamefont{R.}~\bibnamefont{{Reyes}}}, \bibinfo
  {author} {\bibfnamefont{R.}~\bibnamefont{{Mandelbaum}}}, \bibinfo {author}
  {\bibfnamefont{U.}~\bibnamefont{{Seljak}}}, \bibinfo {author}
  {\bibfnamefont{T.}~\bibnamefont{{Baldauf}}}, \bibinfo {author}
  {\bibfnamefont{J.~E.}\ \bibnamefont{{Gunn}}}, \bibinfo {author}
  {\bibfnamefont{L.}~\bibnamefont{{Lombriser}}},\ and\ \bibinfo {author}
  {\bibfnamefont{R.~E.}\ \bibnamefont{{Smith}}},\ }%
  \bibfield{journal}{%
  \Doi{10.1038/nature08857}{\bibinfo {journal} {Nature}}\ }%
  \textbf{\bibinfo {volume} {464}},\ \bibinfo {pages} {256} (\bibinfo {month}
  {Mar.}\ \bibinfo {year} {2010}),\
  \Eprint{http://arxiv.org/abs/1003.2185}{arXiv:1003.2185 [astro-ph.CO]}%
  \bibAnnoteFile{NoStop}{2010Natur.464..256R}%
\bibitem{arfken+weber}%
  \BibitemOpen
  \bibfield{author}{%
  \bibinfo {author} {\bibfnamefont{G.~B.}\ \bibnamefont{{Arfken}}}\ and\
  \bibinfo {author} {\bibfnamefont{H.~J.}\ \bibnamefont{{Weber}}},\ }%
  \emph{\bibinfo {title} {{Mathematical Methods for Physicists}}},\ \bibinfo
  {edition} {6th}\ ed.\ (\bibinfo {publisher} {Elsevier Academic Press},\
  \bibinfo {year} {2005})\ ISBN \bibinfo {isbn} {0-12-088584-0}%
  \bibAnnoteFile{NoStop}{arfken+weber}%
\end{thebibliography}%


\end{document}